\documentclass[prc,preprint,showpacs,showkeys,
tightenlines,
        amsfonts,nofootinbib]{revtex4}

\usepackage{graphicx}  %
\usepackage{bm}  %
\usepackage{amssymb}
\usepackage{dcolumn}

\begin{document}

%


\newcommand{\beq}{\begin{equation}}
\newcommand{\eeq}{\end{equation}}
\newcommand{\bea}{\begin{eqnarray}}
\newcommand{\eea}{\end{eqnarray}}
\newcommand{\ben}{\begin{eqnarray*}}
\newcommand{\een}{\end{eqnarray*}}

\newcommand{\simlt}{\stackrel{<}{{}_\sim}}
\newcommand{\simgt}{\stackrel{>}{{}_\sim}}
\newcommand{\sing}{$^1\!S_0$ }
\newcommand{\btau}{\mbox{\boldmath$\tau$}}
\newcommand{\bsig}{\mbox{\boldmath$\sigma$}}

\newcommand{\dt}{\partial_t}

\newcommand{\kf}{k_{\rm F}}
\newcommand{\wt}{\widetilde}
\newcommand{\kt}{\widetilde k}
\newcommand{\pt}{\widetilde p}
\newcommand{\qt}{\widetilde q}
\newcommand{\wh}{\widehat}
\newcommand{\dens}{\rho}
\newcommand{\edens}{{\cal E}}
\newcommand{\order}[1]{{\cal O}(#1)}

\newcommand{\psihat}{\widehat\psi}
\newcommand{\dagphan}{{\phantom{\dagger}}}
\newcommand{\kvec}{{\bf k}}
\newcommand{\kpvec}{{\bf k}'}
\newcommand{\ak}{a^\dagphan_\kvec}
\newcommand{\akdag}{a^\dagger_\kvec}
\newcommand{\akv}[1]{a^\dagphan_{\kvec_{#1}}}
\newcommand{\akdagv}[1]{a^\dagger_{\kvec_{#1}}}
\newcommand{\akp}{a^\dagphan_{\kvec'}}
\newcommand{\akpdag}{a^\dagger_{\kvec'}}
\newcommand{\akpv}[1]{a^\dagphan_{\kvec'_{#1}}}
\newcommand{\akpdagv}[1]{a^\dagger_{\kvec'_{#1}}}

\def\vec#1{{\bf #1}}

\newcommand{\nab}{\overrightarrow{\nabla}}
\newcommand{\nabsq}{\overrightarrow{\nabla}^{2}\!}
\newcommand{\nabl}{\overleftarrow{\nabla}}
\newcommand{\galnab}{\tensor{\nabla}}
\newcommand{\psid}{{\psi^\dagger}}
\newcommand{\psidal}{{\psi^\dagger_\alpha}}
\newcommand{\psidbe}{{\psi^\dagger_\beta}}
\newcommand{\idt}{{i\partial_t}}
\newcommand{\Sthree}{{\delta_{11'}(\delta_{22'}\delta_{33'}%
        -\delta_{23'}\delta_{32'})%
        +\delta_{12'}(\delta_{23'}\delta_{31'}-\delta_{21'}\delta_{33'})%
        +\delta_{13'}(\delta_{21'}\delta_{32'}-\delta_{22'}\delta_{31'})}}
\newcommand{\Stwo}{{\delta_{11'}\delta_{22'}-\delta_{12'}\delta_{21'}}}
\newcommand{\Left}{{\cal L}}
\newcommand{\Tr}{{\rm Tr}}

\newcommand{\h}{\hfil}
\newcommand{\be}{\begin{enumerate}}
\newcommand{\ee}{\end{enumerate}}
\newcommand{\I}{\item}   

\newcommand{\density}{\rho}

\newcommand{\thyp}{\mbox{---}}

\newcommand{\Jks}{J_{\ks}}
\newcommand{\Jzero}{\Jks}
\newcommand{\Jdensityzero}{J_\density^0}

\newcommand{\ks}{{\rm ks}}
\newcommand{\Seq}{Schr\"odinger\ equation }
\newcommand{\yvec}{{\bf y}}
\newcommand{\ve}{V_{eff}}
\newcommand{\densityJzero}{\density^0_J}
\newcommand{\Dfunct}{{D^{-1}}}
\newcommand{\drv}[2]{{\mbox{$\partial$} #1\over \mbox{$\partial$} #2}}
\newcommand{\drvs}[2]{{\mbox{$\partial^2$} #1\over \mbox{$\partial$} #2 \mbox{$^2$}}}
\newcommand{\drvt}[2]{{\partial^3 #1\over \partial #2 ^3}}
\newcommand{\til}[1]{{\widetilde #1}}
\newcommand{\dthreex}{d^3\xvec}
\newcommand{\dthreey}{d^3\yvec}

\newcommand{\efermi}{\varepsilon_{{\scriptscriptstyle \rm F}}}
\newcommand{\eHF}{\wt\varepsilon}
\newcommand{\eKS}{e}
\newcommand{\ekJ}{e_\kvec^J}
\newcommand{\epsk}{\varepsilon_\kvec}
\newcommand{\epsKS}{\varepsilon}
\newcommand{\Eq}[1]{Eq.~(\ref{#1})}

\newcommand{\Fi}[1]{\mbox{$F_{#1}$}}
\newcommand{\fq}{f_{\qvec}}

\newcommand{\Gammaalt}{\overline\Gamma}
\newcommand{\Gammaks}{\Gamma_{\ks}}
\newcommand{\Gamint}{\widetilde{\Gamma}_{\rm int}}
\newcommand{\GKS}{G_{\ks}}
\newcommand{\grad}{{\bm{\nabla}}}   
\newcommand{\greenKS}{{G}_{\ks}}
\newcommand{\intint}{\int\!\!\int}

\newcommand{\kfermi}{k_{{\scriptscriptstyle \rm F}}}   
\newcommand{\kJzero}{k_J}

\renewcommand{\l}{\lambda}

\newcommand{\MeV}{\mbox{\,MeV}}
\newcommand{\mi}[1]{\mbox{$\mu_{#1}$}}

\newcommand{\Oi}[1]{\mbox{$\Omega_{#1}$}}

\newcommand{\phibar}{\overline\phi}
\newcommand{\phidagger}{\phi^\dagger}
\newcommand{\phistar}{\phi^\ast}
\newcommand{\psibar}{\overline\psi}
\newcommand{\psidagger}{\psi^\dagger}
\newcommand{\qvec}{\vector{\rho}}

\newcommand{\tr}{{\rm tr\,}}

\newcommand{\Ulong}{U_{L}}

\renewcommand{\vector}[1]{{\bf #1}}
\newcommand{\vext}{v_{\rm ext}}   
\newcommand{\Vlong}{V_{L}}

\newcommand{\Wzero}{W_0}
\newcommand{\Wks}{W_{\ks}}
\newcommand{\xvec}{\vector{\bf{x}}}
\newcommand{\zvec}{\vector{z}}
\newcommand{\rvec}{\vector{r}}

\newcommand{\ts}{\textstyle}
%
%
%
%


%
\title{The Kinetic Energy Density in Kohn-Sham\\ Density Functional Theory}
\author{Anirban Bhattacharyya}\email{anirban@mps.ohio-state.edu}
\author{R.J. Furnstahl}\email{furnstahl.1@osu.edu}

\affiliation{Department of Physics,
         The Ohio State University, Columbus, OH\ 43210}

%
\date{August, 2004}

\begin{abstract}
%
This work continues a program to systematically generalize 
the Skyrme Hartree-Fock method for medium and heavy nuclei by applying
effective field theory (EFT) methods to Kohn-Sham density 
functional theory (DFT). 
When conventional Kohn-Sham DFT for Coulomb systems is extended beyond
the local density approximation, the kinetic energy
density $\tau$ is sometimes
included in energy functionals in addition to the fermion density.
 However, a local (semi-classical) expansion of $\tau$ 
 is used to write the energy as a functional of the density 
alone, in contrast to the Skyrme approach. 
The difference is manifested in different 
single-particle equations, which in the Skyrme case include a spatially
varying effective mass. Here we show how to generalize 
the EFT framework for DFT derived previously
to reconcile these approaches. A dilute gas of 
fermions with short-range interactions confined by an external potential 
serves as a model system for comparisons and for testing power-counting
estimates of new contributions to the energy functional.
\end{abstract}

\smallskip
\maketitle


\section{Introduction}
\label{sec:introduction}

Density functional theory (DFT) is widely used in many-body applications
with Coulomb interactions because
it provides a 
useful balance between accuracy and computational cost, allowing large systems 
to be treated in a simple self-consistent manner. 
In DFT, the particle density  
$\rho{(\xvec)}$ plays a central role rather than the many-body wave 
function. 
DFT has the generality to deal with any interaction but  
has had little explicit impact on
nuclear structure phenomenology so far (see, however
Refs.~\cite{PETKOV91,HOFMANN98,Fayans:2001fe,SOUBBOTIN03}), although
the Skyrme-Hartree-Fock formalism is often considered to be a form of DFT
\cite{BRACK85}. 
In previous work, effective field theory (EFT) methods
were applied in a DFT framework, with
the ultimate goal of calculating bulk observables 
for medium to heavy nuclei in 
a systematic fashion.
The present work takes another step toward this goal by generalizing
the EFT framework for DFT to include the kinetic energy density in the
same way it appears in the Skyrme approach.

Kohn-Sham (KS) DFT maps an interacting many-body system to a 
much easier-to-solve non-interacting system. 
Hohenberg and Kohn proved that the ground-state density $\rho{(\xvec)}$ 
of a bound system of 
interacting particles in some external potential $v(\xvec)$ determines this 
potential 
uniquely up to an additive constant~\cite{HK64}.
They showed that there exists 
an energy 
functional $E[\rho]$ that can be decomposed as 
\beq
   E[\rho(\xvec)] = F_{\rm HK}[\rho(\xvec)] 
      + \int\! \dthreex\, v(\xvec)\rho(\xvec) \ ,
      \label{eq:HKuniv}
\eeq
where the functional $F_{\rm HK}[\rho]$ is called the HK free energy,  
and is universal in the sense that it has no {\em explicit} dependence
on the potential $v(\xvec)$.
A variational principle,
formulated in terms of trial densities rather than 
trial wavefunctions,
ensures that the functional $E[\rho]$ is
a minimum equal to the ground-state energy when evaluated at the exact
ground-state density. 

The free-energy functional
$F_{\rm HK}[\rho]$ can be further decomposed into a non-interacting kinetic
energy $T_s[\rho]$, the Hartree term $E_H[\rho]$, and the 
exchange-correlation energy 
$E_{xc}[\rho]$ \cite{ARGAMAN00}:
\beq
   F_{\rm HK}[\rho] = T_s[\rho] + E_H[\rho] + E_{xc}[\rho] \ .
\eeq
Kohn-Sham DFT~\cite{KOHN65,PARR89,DREIZLER90,KOHN99,ARGAMAN00} 
allows $T_s[\rho]$ to be calculated exactly [see \Eq{T0}],
which efficiently treats a leading source of non-locality in the energy
functional. 
Given the Hartree and exchange-correlation functionals,
the solution of the interacting problem at zero temperature
reduces to solving a 
single-particle Schr\"odinger equation, 
 \beq
  \bigl[ -\frac{{\nabla}^2}{2M}  +  v_s(\xvec)
  \bigr]\, \psi_\beta({\bf x}) = \varepsilon_\beta 
              \psi_\beta({\bf x})
  \, ,
  \label{eq:schrodKS}
 \eeq
for the lowest $A$ 
orbitals $\psi_\beta(\xvec)$ (including degeneracies),%
\footnote{This is only true in the absence of pairing.  See
Ref.~\cite{FURNSTAHL04} for a discussion 
on generalizing DFT/EFT to accommodate pairing
correlations.}
where the {\em effective} local external 
potential $v_{s}(\xvec)$ is given by
\beq
v_s(\xvec) = v(\xvec) + \frac{\delta E_{H}[\rho]}{\delta \rho{(\xvec)}} + 
\frac{\delta E_{xc}[\rho]}{\delta \rho{(\xvec)}}\, .\label{eq:vks}
\eeq  
The wave functions for the occupied states in this 
fictitious external potential $v_s(\xvec)$ generate the same 
physical density $\rho{(\xvec)}$ through $\rho(\xvec) \equiv \sum_\beta
|\psi_\beta(\xvec)|^2$ 
as that of the original, fully interacting system in the external 
potential $v(\xvec)$.

The practical problem of DFT is finding a useful explicit
expression for $E_{xc}[\rho]$
\cite{DREIZLER90,PARR89,ARGAMAN00,KOHN99}.
For Coulomb systems, the conventional procedure is to approximate
$E_{xc}$  in the local density approximation (LDA) by taking the 
exchange-correlation energy 
density at each point in the system to be equal to its value 
for a {\em uniform} interacting 
system at the local density
(which is calculated numerically) \cite{ARGAMAN00}, 
and then to include semi-phenomenological gradient 
corrections \cite{PERDEWWANG92,ARGAMAN00}. 
These gradient corrections have become steadily more sophisticated.
The initial attempts at gradient corrections violated a sum rule 
\cite{KOHN99,ARGAMAN00}, which was fixed
by the generalized gradient expansion approximation (GGA), 
where spatial variations 
of $\rho({\xvec})$ are constrained by construction
to conform with the sum rule \cite{PERDEW96}.  
To further improve the exchange-correlation functional, 
the semi-local kinetic 
energy density $\tau{(\xvec)} \equiv
\sum_\beta|\bm{\nabla}\psi_\beta(\xvec)|^2$ was 
built into the GGA formalism to construct 
meta-generalized gradient approximations (Meta-GGA) \cite{PERDEW99}. 
However, in practice, the kinetic energy density is replaced by its 
local expansion in terms of the density $\rho({\xvec})$ and its 
gradients \cite{DREIZLER90} so that 
the energy is treated as a functional of the density alone.
The Kohn-Sham single-particle equation still takes the form of
Eq.~(\ref{eq:schrodKS}). 

The Skyrme-Hartree-Fock approach to nuclei is also based on
energy functionals and single-particle equations,
which suggests a link between the  traditional
DFT and nuclear mean-field approaches
\cite{BRACK85,SCHMID95,SCHMID95a}. 
In calculations with the usual density-dependent Skyrme force,
the energy density for spherical, 
even-even $N = Z$ nuclei takes the form of a local
expansion in density \cite{VB72, RINGSCHUCK}, 
\bea
\edens_{SK}{(\xvec)} &=& \frac{1}{2M} \tau{(\xvec)} +
\frac{3}{8}t_0 {[\rho{(\xvec)}]}^2 + 
\frac{1}{16}t_3{[\rho{(\xvec)}]}^{2+\alpha} + 
\frac{1}{16}(3t_1 + 5t_2)\rho{(\xvec)}\,\tau{(\xvec)} 
\nonumber \\
     & &
+ \frac{1}{64}(9t_1 - 5t_2)|\bm{\nabla} \rho(\xvec)|^2 
 - \frac{3}{4}W_0 \rho{(\xvec)}\bm{\nabla}\cdot \bm{J}(\xvec) + 
\frac{1}{32}(t_1 - t_2)[{\bm{J}}(\xvec)]^2 \, ,\label{eq:SkyrmeEdens}
\eea
(see Ref.~\cite{JACEK95} for a more general treatment).
The density $\rho$, kinetic density $\tau$, and the spin-orbit
density $\bm{J}$ are expressed as sums over single-particle
orbitals $\psi_\beta(\xvec)$: 
\beq
  \rho(\xvec) \equiv \sum_\beta |\psi_\beta(\xvec)|^2\,,
  \qquad
  \tau(\xvec) \equiv \sum_\beta |\bm{\nabla}\psi_\beta(\xvec)|^2
  \,, \qquad
  \bm{J}(\xvec) \equiv \sum_\beta \psi^\dagger_\beta(\xvec)
    (\bm{\nabla\times\sigma})\psi_\beta(\xvec)
  \,,
  \label{eq:skyrmeeqs}
\eeq
where the sums are over occupied states.
The $t_i$'s, $W_0$, and $\alpha$ are generally obtained from numerical fits to
experimental data.
Varying the energy with
respect to the wavefunctions leads to a Schr\"odinger-type equation 
with a position-dependent mass term \cite{VB72, RINGSCHUCK}:
\beq
\Bigl( - \bm{\nabla} \frac{1}{2M^*(\xvec)} \bm{\nabla} + U(\xvec) + 
 W(\xvec) 
 \Bigr)\,
\psi_\beta(\xvec) =
\varepsilon_\beta\,\psi_\beta(\xvec)\, ,\label{eq:SkyrmeEq}
\eeq
where $M^*(\xvec)$ is given by
\beq
\frac{1}{2M^{*}(\xvec)}
 =\frac{1}{2M} + \left[{3\over {16}}\,t_1
     +{5\over {16}}\,t_2\right]\,\rho(\xvec)\,
     ,\label{eq:MassEq}
\eeq
and $W(\xvec)$ is a spin-orbit potential (see Ref.~\cite{JACEK95} for
details).
The appearance of $M^*(\xvec)$ and $W(\xvec)$ 
are a consequence of not expanding $\tau$ and
$\bm{J}$ in terms of $\rho$.

The Skyrme approach has had many phenomenological successes over the last
30 years and, generalized to include the effects of pairing
correlations, continues to be 
a major tool for analyzing the nuclear structure of medium and heavy 
nuclei \cite{BROWN98,Dobaczewski:2001ed,BENDER2003,Stoitsov:2003pd}.
The form of the Skyrme interaction was originally motivated as an expansion
of an effective interaction (G-matrix) in the medium \cite{SKYRME}.
Negele and Vautherin
made the connection concrete with the density matrix expansion
method \cite{DME}, but there has been little further development since
their work.
Many unresolved questions remain, which become acute as one attempts to
reliably extrapolate away from well-calibrated nuclei and to connect to modern
treatments of the few-nucleon problem. 
Why are certain terms included in the energy functional
and not others?  What is the expansion
parameter(s)?  How can we estimate uncertainties in the predictions?
Are long-range effects included adequately?  Can we
really include correlations beyond mean-field, as implied by the DFT
formalism?
We turn to effective field theory to address these questions, using a DFT
rather than a G-matrix approach as in Ref.~\cite{Duguet:2002xn}.

Effective field theory (EFT) promises a  model-independent framework 
for analyzing 
low-energy phenomena with reliable error estimates 
\cite{EFT98,EFT99,LEPAGE89,Birareview,BEANE99}. 
In Ref.~\cite{PUG02}, an effective action framework was
used to merge effective field theory with
DFT for a dilute gas of identical fermions confined in an external potential
with short-range, spin-independent
interactions (see also Refs.~\cite{Polonyi:2001uc,Schwenk:2004hm}).  
The calculations in Ref.~\cite{PUG02} used the results from the EFT treatment
of a uniform system in Ref.~\cite{HAMMER00} in the local density
approximation.
In the present work, we extend the formalism to include the kinetic
energy density as a functional variable, 
leading to Kohn-Sham equations with position-dependent
$M^*$'s
(see Ref.~\cite{AUREL95} for an earlier discussion of such an extension
to Kohn-Sham DFT).
We take a (small) step beyond the LDA by evaluating the full $\tau$ dependence
in the Hartree-Fock diagrams with two-derivative vertices, 
which leads to an energy expression
similar to the standard Skyrme energy (excepting the spin-orbit parts).

The plan of the paper is as follows.
In Sect.~\ref{sect:dilute}, 
we extend the EFT/DFT
construction for a dilute Fermi system
to include a source coupled to the kinetic energy density operator.
A double Legendre transformation, carried out via the inversion
method, yields an energy functional of $\rho$ and $\tau$, and Kohn-Sham
equations with $M^*(\xvec)$.
In Sect.~\ref{sect:results}, we illustrate the formalism with
numerical calculations
for a dilute Fermi system in a trap, including power-counting estimates.
Section~\ref{sect:summary} is a summary.
For the bulk of the paper, we will restrict our discussion to
spin-independent interactions but in Sec.~\ref{sec:skyrme}
the extension to more general forces is discussed.


\section{EFT/DFT with the Kinetic Energy Density}
\label{sect:dilute}

A dilute Fermi system with short-range interactions is an ideal test 
laboratory for effective field
theory at finite density, but it is also directly relevant for
comparison to Skyrme functionals.
We describe such a system using
a general local Lagrangian for a nonrelativistic fermion
field with spin-independent, short-range interactions
that is invariant under Galilean, parity, and time-reversal
transformations \cite{BEANE99,HAMMER00}:
\bea
  {\cal L}  &=&
       \psi^\dagger \biggl[i\partial_t + \frac{\nab^{\,2}}{2M}\biggr]
                 \psi - \frac{C_0}{2}(\psi^\dagger \psi)^2
            + \frac{C_2}{16}\Bigl[ (\psi\psi)^\dagger
                                  (\psi\galnab^2\psi)+\mbox{ h.c.}
                             \Bigr]
  \nonumber \\[5pt]
   & & \null +
         \frac{C_2'}{8} (\psi \galnab \psi)^\dagger \cdot
             (\psi\galnab \psi)
+  \cdots \ ,
  \label{eq:lag}                                                   
\eea
where $\galnab=\overleftarrow{\nabla}-\nab$ is the Galilean invariant
derivative and h.c.\ denotes the Hermitian conjugate.
The terms proportional to $C_2$ and $C_2'$ contribute to $s$-wave and
$p$-wave scattering, respectively, 
while the dots represent terms with more derivatives and/or more
fields.
To describe trapped fermions,
we add to the Lagrangian a  
term for an external confining potential $v(\xvec)$  
coupled to the density 
operator $v(\xvec)\psi^\dagger\psi$ \cite{PUG02}.
For the numerical calculations, we take the potential 
to be an isotropic harmonic confining potential,
\beq
v(\xvec)={1\over 2} m\,\omega^2 \, |\xvec|^2 \ ,
 \label{eq:vext}
\eeq
although the discussion holds for a general non-vanishing
external potential.
This Lagrangian can be simply generalized to include spin-dependent
interactions, as described in Sec.~\ref{sec:skyrme}, although the 
complete set of terms grows rapidly with the number of derivatives. 

\subsection{Effective Action and the Inversion Method}
\label{sect:effact}

We introduce a generating functional in the path integral formulation
with ${\cal L}$ from Eqs.~(\ref{eq:lag}) and (\ref{eq:vext})
supplemented by
 c-number
sources $J(x)$ and $\eta(x)$, coupled 
to the composite density operator and to the
kinetic energy density operator, respectively,   
\beq
    Z[J,\eta] = e^{iW[J,\eta]}
      = \int\! D\psi D\psi^\dag \ e^{i\int\! d^4x\ [{\cal L}\,+\,   J(x)
    \psi^\dag(x) \psi(x)\,+\,
    \eta(x)\bm{\nabla}\psi^\dag(x)\cdot\bm{\nabla}\psi(x)]}
    \ .
   \label{partfunc1}
\eeq
For simplicity, normalization factors are considered to be implicit
in the functional integration measure \cite{FUKUDA94,FUKUDA95}.
The fermion density in the presence of the sources is
\beq
\rho(x)\equiv \langle \psi^\dag(x)\psi (x)\rangle_{J\,,\,\eta}
= {\delta W[J,\eta]\over \delta J(x)} \ ,
\label{partden1}
\eeq
and the kinetic energy density is given by
\beq
\tau(x)\equiv \langle \bm{\nabla}\psi^\dag(x)\bm{\cdot\nabla}\psi (x)\rangle_{J\,,\,\eta}
= {\delta W[J,\eta]\over \delta \eta(x)} \ ,
\label{partkin1}
\eeq
where functional derivatives are taken keeping the other
source fixed. 
The effective action is defined  
through the functional Legendre transformation 
\beq
\Gamma[\rho,\tau]  = 
  W[J,\eta]- \int\! d^4x\, J(x)\rho(x) - \int\! d^4x\, \eta(x)\tau(x)\   ,
\label{ehk}
\eeq
which implies that $\Gamma$ has no {\em explicit} dependence on
$J$ and $\eta$. 

As in Ref.~\cite{PUG02}, we choose
finite-density boundary conditions that
enforce a given particle number $A$ by hand, so that 
$\Gamma$ and $W$ are functions of $A$
and variations over $\rho(x)$ and $\tau(x)$ conserve $A$.
This is naturally achieved by working with a fixed number of Kohn-Sham
orbitals, as seen below.
By limiting ourselves to time independent sources and densities,   
we can factor out a ubiquitous time factor 
corresponding to the time interval over 
which the source is acting, and write \cite{PUG02}
\beq
\til{\Gamma}[\rho,\tau]\equiv \Gamma[\rho,\tau] 
  \times \left[\int_{-\infty}^{\infty}\, dt\right]^{-1} 
  = - E[\rho,\tau]
   \label{eq:LamE}
  \, .
\eeq
and similarly with $W[J,\eta]$ and the expansions below.
The effective action has extrema at the
possible quantum ground states of the system, and when evaluated at the
minimum is proportional (at zero temperature) to the ground-state
energy \cite{FUKUDA94,FUKUDA95,HU96}.
In particular, Eq.~(\ref{eq:LamE}) defines an energy functional
$E[\rho,\tau]$, which when evaluated with the exact ground-state
density $\rho$ and kinetic energy density $\tau$, is 
equal to the ground-state energy.

If we take functional derivatives of Eq.~(\ref{ehk})
with respect to $J(\xvec)$ and $\eta(\xvec)$, and then apply 
Eqs.~(\ref{partden1}) and (\ref{partkin1}) \cite{PUG02}, 
we obtain two equations that can be written
in matrix form as
\beq
   \int d^3\yvec\,   
    \left (
   \begin{array}{lr}
   \frac{\ts\delta\rho(\yvec)\phantom{\til{\Gamma}}}{\ts\delta J(\xvec)} &  
     \frac{\ts\delta\tau(\yvec)}{\ts\delta J(\xvec)} \\[10pt]
   \frac{\ts\delta\rho(\yvec)\phantom{\til{\Gamma}}}{\ts\delta \eta(\xvec)} &
   \frac{\ts\delta\tau(\yvec)}{\ts\delta \eta(\xvec)}
   \end{array}
   \right ) 
  \,
   \left(
   \begin{array}{l}
     \frac{\ts\delta \til{\Gamma}[\rho,\tau]}{\ts\delta\rho(\yvec)}
      + J(\yvec) \\[10pt]
     \frac{\ts\delta \til{\Gamma}[\rho,\tau]}{\ts\delta\tau(\yvec)}
      + \eta(\yvec)  
   \end{array}
   \right )
   =0     \ .
   \label{eq:matrixa}
\eeq
The invertibility of the transformation from $\{J,\eta\}$ to
$\{\rho,\tau\}$ implies that the matrix in Eq.~(\ref{eq:matrixa}) has no
zero eigenvalues, which means that the elements of the vector vanish
identically for each $\xvec$, so that
\beq
    \frac{\delta\til{\Gamma}[\rho,\tau]}{\delta\rho(\xvec)}
  = -J(\xvec) \ , \qquad
    \frac{\delta\til{\Gamma}[\rho,\tau]}{\delta\tau(\xvec)}
  = -\eta(\xvec)  
  \ .
\label{min1}
\eeq
Thus, the effective action when evaluated at the exact ground
state density and kinetic energy density is an extremum when 
the sources are set
to zero, which corresponds to the original (source-free) system.
The convexity of $\Gamma$ implies that the energy (equal to
minus $\til{\Gamma}[\rho,\tau]$ at the extremum) is a minimum.%
\footnote{In Ref.~\cite{VALIEV97}, a proof is given of the invertibility
of the Legendre transformation for the Euclidean version of the
functions.  The proof extends directly to our Minkowski functions with
any number of sources, as long as they are coupled linearly.}

These properties are analogous to conventional applications of effective
actions where the Legendre transformation is with respect to one of the
fields rather than a composite operator \cite{COLEMAN88,PESKIN95,WEINBERG96}.
A possible complication in the present case would be if new divergences
arose in $W[J,\eta]$, which generally happens when adding a source
coupled to a composite operator \cite{BANKSRABY,COLLINS86}.  
However, that is not the case here \cite{FURNSTAHL04}.  
On the other hand, sources coupled to operators such as $\psi\psi +
\psi^\dagger\psi^\dagger$, which arise when considering pairing, 
\emph{will}
introduce new divergences, including for the kinetic energy density.
This issue will be considered elsewhere \cite{FURNSTAHL04}.

To carry out the inversion and Legendre transformation, we apply the
inversion method of Fukuda et al.\ 
\cite{FUKUDA94,FUKUDA95,VALIEV96,VALIEV97,VALIEV97b,RASAMNY98}.
This method applies to any system that can be characterized by a
hierarchy, which we organize by introducing a parameter $\lambda$
that is ultimately set to unity.
This parameter can label different orders in a coupling constant
expansion (e.g., powers of $e^2$ for the Coulomb interaction),
in a large $N$ expansion, or in an effective field theory expansion.
Here we apply the latter, and  associate powers of $\lambda$
with the orders in the
dilute EFT expansion derived in Refs.~\cite{HAMMER00} and
\cite{PUG02}.
The 
effective action is given a dependence
on $\lambda$ 
\beq
    \til{\Gamma} = \til{\Gamma}[\rho,\tau,\lambda]
    \ ,
\eeq
which is treated as an independent variable.
The Legendre transformation defining $\til{\Gamma}$ follows from 
Eq.~(\ref{ehk}): 
\beq
  \til{\Gamma}[\rho,\tau,\lambda] = \til{W}[J,\eta,\lambda] - 
  \int \! \dthreex \ J (\xvec)\ \rho(\xvec)
- \int \! \dthreex \ \eta (\xvec)\ \tau(\xvec) 
  \ ,
  \label{eq:Gammalam}
\eeq
where $J$ and $\eta$ depend on $\lambda$ as well as being 
functionals of $\rho$ and $\tau$.

\begin{figure}[t]
\centerline{\includegraphics*[width=12cm,angle=0]{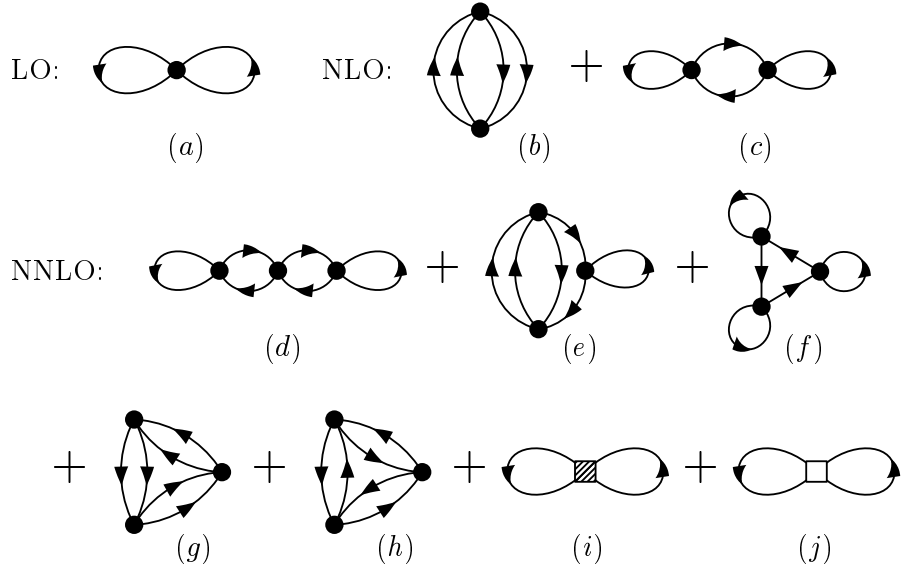}}
\vspace*{-.1in}
\caption{Hugenholtz diagrams for a dilute Fermi gas through
next-to-next-to-leading-order (NNLO) in an EFT expansion.}
\label{fig:hug}
\end{figure}

Now we expand each of the quantities
that depend on $\lambda$ in \Eq{eq:Gammalam} in
a series in $\lambda$ \cite{PUG02},
treating $\rho$ and $\tau$ as order unity,
and substitute the expansion for $J$
and $\eta$ into the expansion for
$\til{W}$. 
Equating terms with
equal powers of $\lambda$ on both sides of \Eq{eq:Gammalam} 
after carrying out a functional Taylor expansion of $\til{W}[J,\eta]$ about
$J_0$ and $\eta_0$ gives 
a series of equations relating the $\til{\Gamma}_l$, $\til{W}_l$,
$J_l$, and $\eta_l$, where the subscript $l$ indicates the power of
$\lambda$.
These equations allow the $\til{\Gamma}_l$'s to be constructed
recursively (see Ref.~\cite{PUG02} for explicit expressions).
Since $\rho$ and $\tau$ are independent of $\lambda$,
the sources for \emph{any} $l$ satisfy
\beq
J_l(\xvec)
     = - \frac{\delta \til{\Gamma}_l[\rho,\tau]}{\delta
     \rho(\xvec)}\ , \qquad
\eta_l(\xvec)
     = - \frac{\delta \til{\Gamma}_l[\rho,\tau]}{\delta
     \tau(\xvec)} 
       \ ,
    \label{eq:sources}
\eeq    
which is just the term-by-term expansion of \Eq{min1}.
All of the $J_l$'s and $\eta_l$'s as
defined here are functionals of $\rho$ and $\tau$, and this
functional dependence will be understood even if not explicitly
shown from now on (and functional derivatives with respect to $\rho$
will have $\tau$ held fixed, and vice versa). 
We identify the $\til{W}_l$'s for $l\geq 1$ in the present case with the
diagrammatic expansion in Fig.~\ref{fig:hug} \cite{HAMMER00,PUG02}.
That is, $\til{W}_1$ is given by diagram (a), $\til{W}_2$ by diagrams
(b) and (c), and $\til{W}_3$ by diagrams (d) through (j).

The zeroth-order equation from Eq.~(\ref{eq:Gammalam}) is 
\beq
  \til{\Gamma_0}[\rho,\tau] = \til{W_0}[J_0,\eta_0] - 
  \int \! \dthreex \ J_0 (\xvec)\ \rho(\xvec)
- \int \! \dthreex \ \eta_0 (\xvec)\ \tau(\xvec) 
       \ .
    \label{eq:Gamma_0} 
\eeq
The corresponding zeroth order expansion of Eqs.~(\ref{partden1})
and (\ref{partkin1}) is
\beq
   \rho(\xvec) = \frac{\delta \til{W}_0[J_0,\eta_0]}{\delta
   J_0(\xvec)} \ , \qquad 
 \tau(\xvec) = \frac{\delta \til{W}_0[J_0,\eta_0]}{\delta
   \eta_0(\xvec)} \ ,   
   \label{eq:KSq}
\eeq
(this also follows by taking functional derivatives of
Eq.~(\ref{eq:Gamma_0}) with respect to $\rho$ and $\tau$ 
and using invertibility again, see \cite{PUG02}).
Note that these are the exact ground-state densities.
Because $\rho$ and $\tau$ are treated as order unity, this is the only
equation in the expansion of Eqs.~(\ref{partden1}) and (\ref{partkin1}) 
in which they appear. 
Thus,
the sources $J_0(\xvec)$ and $\eta_0(\xvec)$ are
\emph{particular} functions that  
generate the expectation values 
$\rho$ and $\tau$ from the {\em noninteracting\/} system 
defined by $\lambda=0$. 
(The existence of $J_0(\xvec)$ and $\eta_0(\xvec)$ 
is the heart of the Kohn-Sham approach.)
The inversion method achieves this end by construction.

The exponent in the non-interacting generating functional $Z_0[J,\eta]$
is quadratic in the fermion fields, 
\bea
&&\int\! d^4x\ \psi^\dagger \biggl[
i\partial_t + \frac{\nabla^{\,2}}{2M} -v(\xvec)+J_0(\xvec)
\biggr]\psi + \int\! d^4x\ \eta_{0}(\xvec)\nabla\psi^\dag\cdot\nabla\psi
\\ \nonumber
&&
=
\int\! d^4x\ \psi^\dagger \biggl[
i\partial_t + \frac{\nabla^{\,2}}{2M} -v(\xvec)+J_0(\xvec) 
-\nabla\cdot \eta_{0}(\xvec)\nabla
\biggr]\psi 
\ ,
\eea
which leads us to define
the Green's
function $\greenKS^0$ of the Kohn-Sham non-interacting system.
This Green's function satisfies
\beq
\left( i \partial_t+ \nabla\cdot 
  \frac{1}{2M^{*}(\xvec)}
\nabla -v(\xvec)+J_0(\xvec)\right) 
\greenKS^0 (\xvec t,\xvec't')=\delta^3(\xvec-\xvec')\delta(t-t')
\label{eq:GKSEq}
\eeq
with finite density boundary conditions \cite{FETTER71} and a
position-dependent effective mass defined by 
\beq
  \frac{1}{2M^{*}(\xvec)} \equiv \frac{1}{2M} - \eta_0(\xvec) 
  \ .
  \label{eq:Mstar}
\eeq
Kohn-Sham orbitals arise as solutions to
\beq
  \bigl[ - \nabla\cdot 
  \frac{1}{2M^{*}(\xvec)}
  \nabla 
  +  v(\xvec) -  J_0(\xvec)
  \bigr]\, \psi_k({\bf x}) = \varepsilon_k \psi_k({\bf x})
  \, ,
 \label{eq:ks1}  
\eeq 
where the index $k$ represents all quantum numbers except for the spin.
Note that Eq.~(\ref{eq:ks1}) is in the form of the Skyrme
single-particle
equation [\Eq{eq:SkyrmeEq}] (without the spin-orbit part). 

The spectral decomposition of $\greenKS^0$ in terms
of the Kohn-Sham orbitals is \cite{PUG02}
\beq
   i\greenKS^0 (\xvec t,\xvec't')=\sum_k \psi_k (\xvec)\,
       \psi_k^* (\xvec')\,  
   e^{-i\varepsilon_k(t-t')}[\theta(t-t')\, 
           \theta(\varepsilon_k-\varepsilon_{\rm F})
   -\theta(t'-t)\, \theta(\varepsilon_{\rm F}-\varepsilon_k)]
  \label{eq:Gks}
  \ .
\eeq
It follows that 
$\til{W}_0[J_0,\eta_0] \propto \Tr\ln(\greenKS^0)^{-1}$
(since it is quadratic it yields a simple determinant),
corresponding to the system
without interactions, and can be written   
explicitly in terms of the 
single-particle Kohn-Sham eigenvalues as \cite{NEGELE88} 
\beq
\til{W}_0[J_0,\eta_0]=  -\nu \sum_{\varepsilon_k \le \varepsilon_{\rm F}} 
   \varepsilon_k\ ,
   \label{eq:Wsum}
\eeq
where $\nu$ is the spin-isospin degeneracy,
as expected for a system without interactions.
Equation~(\ref{eq:KSq}) applied to
Eq.~(\ref{eq:Wsum})
[with the help of Eq.~(\ref{eq:ks1})] 
implies that $\rho(\xvec)$ and $\tau(\xvec)$
follow as in Eq.~(\ref{eq:skyrmeeqs}) from the orbitals \cite{PUG02}.
By using \Eq{eq:Wsum} in \Eq{eq:Gamma_0} and then eliminating
$\varepsilon_k$ using Eq.~(\ref{eq:ks1}), 
the lowest order effective action can be written two ways,
\bea
\til{\Gamma}_0[\rho,\tau] &=&
  -\nu\sum_k^{\rm occ.} \varepsilon_k 
    - \int \! \dthreex \ J_0 (\xvec)\ \rho(\xvec) 
    - \int \! \dthreex \ \eta_0 (\xvec)\ \tau(\xvec)     
\nonumber \\
     &=&
-T_s[\tau]-\int \! \dthreex \ v(\xvec)\ \rho(\xvec)  \ ,
\label{G0}
\eea
where 
\beq
 T_s[\tau] = \nu \sum_k\int \! \dthreex \  
 \psi_k^* (\xvec)\left( -{\nabla^2 \over 2M}\right) \psi_k(\xvec) 
  = 
 {\nu \over 2M}\sum_k\int \! \dthreex \ |\nabla\,\psi_k({\bf x})|^2  
 = {1\over 2M}\int \! \dthreex \ \tau(\xvec)
 \label{T0}
 \eeq
is the total kinetic energy of the KS non-interacting system.

The first-order equation in $\lambda$ from Eq.~(\ref{eq:Gammalam}) is
\bea
   \til{\Gamma}_1[\rho,\tau] &=& \til{W}_1[J_0,\eta_0]
   + \int\!\dthreex\,
      \frac{\delta \til{W}_0[J_0,\eta_0]}{\delta J_0(\xvec)} J_1(\xvec) 
   + \int\!\dthreex\,
      \frac{\delta \til{W}_0[J_0,\eta_0]}{\delta \eta_0(\xvec)} \eta_1(\xvec)
      \nonumber \\[5pt]
      &&
   \qquad \null - \int \! \dthreex \ J_1 (\xvec)\ \rho(\xvec)
   - \int \! \dthreex \ \eta_1 (\xvec)\ \tau(\xvec) 
    \nonumber \\ &=&
     \til{W}_1[J_0,\eta_0]
    \ .
    \label{eq:Gamma_1}    
\eea
The complete cancellation of the $J_1$ and $\eta_1$ terms from applying
\Eq{eq:KSq} occurs for the $J_l$ and $\eta_l$ terms in the
equation for $\til{\Gamma}_l$ for all $l\geq 1$. 
Thus for a given $l$, we only need $\til{W}_k$'s with $k$ less than
or equal to $l$, and $J_k$'s and $\eta_k$'s with $k$ smaller
than $l$.
The $\til{W}_k$ functionals are constructed using
conventional Feynman rules in position space with factors 
as in Ref.~\cite{HAMMER00}, but with fermion lines
representing $i\greenKS^0$.

In particular, the  LO effective action is given by \cite{PUG02}
\beq
   \til{\Gamma}_1[\rho,\tau] = \til{W}_1[J_0,\eta_0] = 
{1\over2}\, \nu\,(\nu-1)\, C_0 \int \dthreex\ 
    \greenKS^0 (x,x^+)
  \, \greenKS^0 (x,x^+)  \ .
\label{eq:W1}   
\eeq
The density can be directly expressed in
terms of the Kohn-Sham Green's function with equal arguments as
\beq
  \rho(\xvec) = -i\nu\,\greenKS^0(x,x^+)  \ ,
  \label{eq:denG}
\eeq 
so that we have
\beq 
\til{\Gamma}_1[\rho,\tau]= -{1\over2} {(\nu-1)\over \nu} \ C_0 \int \dthreex 
   \ |\rho (\xvec)|^2 \equiv \til{\Gamma}_1[\rho] \ ,
\label{eq:G1r}
\eeq
which is minus the Hartree-Fock energy.
Using Eq.~(\ref{eq:sources}), we obtain
\beq
J_1(\xvec)=  {C_0\,(\nu-1) \over \nu } \ \rho (\xvec)\ , \qquad\quad 
\eta_1(\xvec)= 0
\ .
\label{eq:J1r}
\eeq

After canceling the $J_2$ and $\eta_2$ terms as advertised above,
the second-order effective action is given by
  \bea
   \til{\Gamma}_2[\rho,\tau] &=& \til{W}_2[J_0,\eta_0]
   + \int\!\dthreex\,
   \frac{\delta \til{W}_1[J_0,\eta_0]}{\delta J_0(\xvec)}
   J_1(\xvec) + \int\!\dthreex\,\frac{\delta \til{W}_1[J_0,\eta_0]}{\delta \eta_0(\xvec)}
   \eta_1(\xvec)
      \nonumber \\[5pt]
      &+& 
   \frac{1}{2}\int\!\dthreex\,\dthreey\,
      \frac{\delta^2 \til{W}_0[J_0,\eta_0]}{\delta J_0(\xvec)\,\delta J_0(\yvec)}
      J_1(\xvec)\, J_1(\yvec) + 
    \frac{1}{2}\int\!\dthreex\,\dthreey\,
      \frac{\delta^2 \til{W}_0[J_0,\eta_0]}{\delta
      \eta_0(\xvec)\,\delta \eta_0(\yvec)}
      \eta_1(\xvec)\, \eta_1(\yvec) 
      \nonumber \\[5pt]
      &+&  
      \int\!\dthreex\,\dthreey\,
      \frac{\delta^2 \til{W}_0[J_0,\eta_0]}{\delta J_0(\xvec)\,\delta
      \eta_0(\yvec)}
      J_1(\xvec)\, \eta_1(\yvec)       
   \ .
     \label{eq:G2}
  \eea
${W}_2[J_0,\eta_0]$ is calculated from the graphs Figs.~\ref{fig:hug}(b)
and (c):
\bea
{W}_2[J_0,\eta_0]&=& i\nu (\nu-1){C_0^2 \over 4} \ \int\! d^4x\  d^4 y\  \greenKS^0 (x,y)\greenKS^0 (x,y)
\greenKS^0(y,x)\greenKS^0(y,x)\nonumber \\
&& \hspace*{3mm} \null - i\nu(\nu-1)^2\ {C_0^2 \over 2} \int\! d^4x \ d^4 y \ 
\greenKS^0 (x,x^+)\greenKS^0 (x,y) \greenKS^0(y,x)\greenKS^0(y,y^+)\ .
\label{eq:W2}
\eea
The other terms in $\til{\Gamma}_2$ completely cancel against
the ``anomalous'' graph of
Fig.~\ref{fig:hug}(c) so that%
\footnote{
This complete
cancellation does \emph{not} occur for long-range forces, or if the
zero-range delta functions at the $C_0$ vertices
are regulated by a cutoff rather than by
dimensional regularization, as used here.}
\beq
{\Gamma}_2[\rho,\tau]= i\nu (\nu-1)\,{C_0^2 \over 4} \int\! d^4x\  d^4 y\  
\greenKS^0 (x,y)\greenKS^0 (x,y) \greenKS^0(y,x)\greenKS^0(y,x) \, ,
  \label{eq:tilgam2}
\eeq
which is equal to the contribution made by
the \lq\lq beachball" diagram [Fig.~\ref{fig:hug}(b)] to the
energy [up to the time factor of \Eq{eq:LamE}].
This cancellation is proven exactly as in Ref.~\cite{PUG02} after
generalizing to the matrix notation of \Eq{eq:matrixa}.
Calculation of the third-order effective action in the inversion
method similarly leads to cancellation of the \lq\lq anomalous" 
graphs in $\til{W}_3$ given by Figs.~\ref{fig:hug}(d), (e), and (f), 
leaving only Fig.~\ref{fig:hug}(g) through (j) as contributors.  
All higher orders in $\til{\Gamma}[\rho,\tau,\lambda]$ are determined 
in a similar manner, as described in Refs.~\cite{VALIEV97,PUG02}. 
In order to solve for the orbitals in
\Eq{eq:ks1} and to calculate the energy, we need 
expressions for $J_0(\xvec)$ and $\eta_0(\xvec)$.
Since $J(\xvec) = \eta(\xvec) = 0$ in the ground state,
Eq.~(\ref{min1}) becomes
a variational principle that, together with \Eq{eq:sources},
yields self-consistent expressions for $J_0$ and $\eta_0$ \cite{PUG02}:
\bea
  J_0(\xvec)\Bigr|_{{\rm gs}} &=& 
    -\sum_{l\geq 1} J_l(\xvec)\Bigr|_{{\rm gs}} = 
    \left.\frac{\delta \Gamint[\rho,\tau]}{\delta \rho(\xvec)}
  \right|_{{\rm gs}} \, , \label{eq:finala}
  \\
\eta_0(\xvec)\Bigr|_{{\rm gs}} &=& 
    -\sum_{l\geq 1} \eta_l(\xvec)\Bigr|_{{\rm gs}} = 
    \left.\frac{\delta \Gamint[\rho,\tau]}{\delta \tau(\xvec)}
  \right|_{{\rm gs}}  
    \ ,
    \label{eq:final}
\eea
where $\Gamint[\rho,\tau]$ is the interaction effective action 
\beq
  \Gamint[\rho,\tau] \equiv \sum_{l\geq 1} \til{\Gamma}_l[\rho,\tau] \ ,
  \label{eq:gamint}
\eeq
and the subscript \lq\lq$\rm gs$" refers to the ground-state. 
(Note that at this stage $\lambda = 1$, and Eqs.~(\ref{eq:finala})
and (\ref{eq:final}) mix all orders of the original inversion-method
expansion into the
Kohn-Sham potentials $J_0$ and $\eta_0$.)
A given approximation corresponds to truncating Eq.~(\ref{eq:gamint})
at $l_{\rm max}$ and then carrying out the self-consistent calculation.
We refer to $l_{\rm max} = 1$ as leading order, or LO,
$l_{\rm max} = 2$ as next-to-leading order or NLO, and
$l_{\rm max} = 3$ as NNLO.

\subsection{EFT for Dilute Fermi Systems}

The non-interacting energy density at zero temperature
for $A$ particles with spin-degeneracy $\nu$ in volume $V$ can be written as
\beq
\edens_0 = {3 \over 5} \left({\kf^2 \over 2 M}\right) \rho \ ,
\label{eqn:E0}
\eeq
where the density  $\rho$ is
\beq
\rho= \frac{A}{V} =
  \nu \int {d^3k\over (2\pi)^3} \ \theta (\kf-k) = {\nu k_F^3\over 6 \pi^2}\ .
\label{unifden}
\eeq
For a uniform system,
the order-by-order in $\kf a_s$
corrections to Eq.~(\ref{eqn:E0}) due to interactions can be calculated
in the EFT via the Hugenholtz diagrams in Fig.~\ref{fig:hug},
as described in Ref.~\cite{HAMMER00}.   

For a dilute Fermi system,
the coefficients $C_0$, $C_2$, and $C_2'$ can be expressed 
in terms of the effective-range 
parameters by matching to the effective-range expansion for low-energy
fermion-fermion scattering~\cite{HAMMER00}:
\beq
      C_0 = \frac{4\pi a_s}{M},\qquad C_2=C_0 \frac{a_s r_s}{2},
               \quad \mbox{and}\quad
          C_2' = \frac{4\pi a_p^3}{M}
      \label{C2imatch}    \,,
\eeq
where
$a_s$ ($a_p$) is the $s$-wave ($p$-wave) scattering length
and $r_s$ is the $s$-wave effective range, respectively.
If the effective range parameters are all of order the interaction
range, then the EFT is said to be natural.
In a uniform system
at finite density, the mean inter-particle spacing $r_0$ provides a length
scale for comparison;
the
ratio $a_s/r_0 \sim k_F a_s$ 
provides a dimensionless measure of density, where 
$k_{F}$ is the Fermi momentum. 
In the dilute regime, $k_F a_s \ll 1$
serves as an expansion parameter, as realized by the EFT of 
Ref.~\cite{HAMMER00} where the diagrams of
Fig.~\ref{fig:hug} each contributed to precisely one order in the
energy density:
LO or ${\cal E}_1$
is ${\cal O}(\kf^6)$, NLO or ${\cal E}_2$ is ${\cal O}(\kf^7)$, 
and NNLO or ${\cal E}_3$ is ${\cal O}(\kf^8)$.

In extending these results to a finite system, the simplest 
approximation that can 
be invoked is the local density approximation
(LDA).
In Ref.~\cite{PUG02}, the results for a uniform system were used in the
LDA to
evaluate the $\til{\Gamma}_l$ and hence the energy through NNLO.
We simply quote the results here. 
 The LO diagram [Fig.~\ref{fig:hug}(a)] is
the Hartree-Fock contribution, which is purely local.
Thus, it is an exact evaluation and \Eq{eq:G1r} can be used directly
for the energy contribution: 
\beq
E_{HF}[\rho(\xvec)] = \int \! \dthreex\ \edens_1[\rho(\xvec)] 
= {1\over2} {(\nu-1)\over \nu} \ C_0 \int \dthreex 
   \ |\rho (\xvec)|^2
\ .
\label{EHF}
\eeq 
The contributions to the energy from NLO and NNLO diagrams are computed 
in LDA by simply integrating the corresponding
uniform energy densities evaluated at the local density \cite{PUG02}:
\bea
  E^{\rm LDA}_{c}[\rho(\xvec)]
    &=& \int \! \dthreex\,  \left\{ \edens_2(\rho_0)
  + \edens_3(\rho_0) \right\}\,|_{\rho_0\rightarrow \rho(\xvec)}
  \nonumber\\
  &=& 
  b_1\, {a_s^2\over 2M} \int \! \dthreex \, [\rho(\xvec)]^{7/3}
    \nonumber  \\ 
  & &
    \null +
    \left( b_2\, a_s^2 \, r_s +b_3\, a_p^3 +b_4\, a_s^3\right) 
     {1\over 2M} \int \! \dthreex \, [\rho(\xvec)]^{8/3}
     \ ,
     \label{eq:LDAc}
\eea
where the dimensionless $b_k$ are
\bea
b_1 &=& {4\over 35\pi^2}\, (\nu-1) \left({6\pi^2\over \nu}\right)^{4/3} 
    \, (11-2\ln2) \ ,\nonumber\\
b_2 &=& {1\over 10 \pi}\, (\nu-1)\left({6\pi^2\over \nu}\right)^{5/3}
    \nonumber \ ,\\ 
b_3 &=& {1\over 5\pi}\, (\nu+1)\left({6\pi^2\over \nu}\right)^{5/3}
    \nonumber \ ,\\
b_4 &=& \left({6\pi^2\over \nu}\right)^{5/3} 
     \biggl(  0.0755\,(\nu-1)+ 0.0574\,(\nu-1)(\nu-3)
\biggr)
     \ .  \label{eq:bfour}
\eea
The numerical constants in the last line of Eq.~(\ref{eq:bfour}) were
obtained by Monte Carlo integration \cite{HAMMER00}.


%
\subsection{Including $\tau$ in Hartree-Fock Diagrams}

The contributions to $E^{\rm LDA}_{c}[\rho(\xvec)]$ from
the Hartree-Fock graphs 
containing the $C_2$ and $C'_2$ vertices [Figs.~\ref{fig:hug}(i) and
(j)],
which have gradients,
were approximately evaluated in the LDA in \Eq{eq:LDAc} to obtain a
functional of $\rho$ alone.
In the DFT formalism generalized to include the kinetic energy density
$\tau(\xvec)$, however, we observe that they can be evaluated exactly
in terms of $\rho(\xvec)$ and $\tau(\xvec)$ (for closed shells). 
These 
 contributions to $\til{W}_3[J_0,\eta_0]$ can be simply expressed
in terms of gradients acting on
the Kohn-Sham Green's functions [cf.\ \Eq{eq:W1}]:
\bea
    \til{W}_3[J_0,\eta_0] &=& 
  -{1\over 16} \int \dthreex\ \biggl[
  C_2\, \nu\,(\nu-1)\, \left\{
    (\bm{\nabla}_1 - \bm{\nabla}_2)^2 +
    (\bm{\nabla}_3 - \bm{\nabla}_4)^2
  \right\}
  \nonumber \\
  & & \qquad\qquad\qquad
  \null + 2C'_2\, \nu\,(\nu+1)\, \left\{
    (\bm{\nabla}_1 - \bm{\nabla}_2) \cdot
    (\bm{\nabla}_3 - \bm{\nabla}_4)
  \right\}\biggr ]
    \nonumber \\ & & \qquad\qquad\qquad\qquad 
    \greenKS^0 (x_4,x_2^+) \, \greenKS^0 (x_3,x_1^+)
    \biggr|_{x_1=x_2=x_3=x_4=x} 
   \ + \cdots
\label{eq:W3}   
\eea
Equation~(\ref{eq:W3}) is evaluated by carrying out the gradients and
then setting all of the $x_i$ equal to $x$.
For this purpose, the replacement
\beq
  \greenKS^0(x,x') \longrightarrow i\sum_k^{\rm occ.}
  \psi_k(\xvec)\psi^*_k(\xvec')
\eeq
can be made and,  since in the end we integrate over $\xvec$, we can
partially integrate any term.
It is not difficult to write the resulting expressions in terms
of $\rho(\xvec)$, $\tau(\xvec)$, and a current density $\bm{j}(\xvec)$,
\beq
  \bm{j}(\xvec) = \frac{i}{2}\nu
    \sum_k^{\rm occ.}\left[
      \bm{\nabla}\psi^*_k(\xvec)\psi_k(\xvec)
      - \psi^*_k(\xvec)\bm{\nabla}\psi_k(\xvec)
    \right] \ .
\eeq 
The current density vanishes when summed over closed shells, leaving
simple expressions for the Hartree-Fock energy functionals:
\beq
  E_{C_2}[\rho(\xvec),\tau(\xvec)]
    = {B_2\, a_s^2\, r_s\over 2M}  \int \! \dthreex \, [\rho(\xvec)\tau(\xvec) 
    + \frac{3}{4}\left( \bm{\nabla}\rho \right)^2] \ , \label{eq:Ec2}
\eeq
\beq
  E_{C^{'}_{2}}[\rho(\xvec),\tau(\xvec)]
    = {B_3\, a_p^3\over 2M}  \int \! \dthreex \, [\rho(\xvec)\tau(\xvec) 
    - \frac{1}{4}\left( \bm{\nabla}\rho \right)^2] \ . \label{eq:Ec2prime}
\eeq
The dimensionless constants $B_2$ and $B_3$ are given by
\beq
B_2 = \pi\, {(\nu-1)\over \nu}  \ , \qquad
B_3 =2\,\pi\, {(\nu+1)\over \nu}  \ .  \label{eq:B2B3}
\eeq

The total contribution to the energy from NLO and NNLO diagrams is
\bea
E_{c}[\rho(\xvec),\tau(\xvec)] &=& 
{b_1\,a_s^2\over 2M} \int \! \dthreex \, [\rho(\xvec)]^{7/3}
 +E_{C_{2}}[\rho(\xvec),\tau(\xvec)]
 \nonumber \\
     & &
 +\,E_{C^{'}_{2}}[\rho(\xvec),\tau(\xvec)]
 +\,{b_4\, a_s^3 \over 2M} \int \! \dthreex \, [\rho(\xvec)]^{8/3}
 \label{eq:Correlation}    \ .
\eea
Since this functional now has the semi-local $\tau(\xvec)$ as one of its 
ingredients, it represents a step beyond LDA, even though  
the contributions from Figs.~\ref{fig:hug}(b), 
(g), and (h) are still
evaluated with the LDA prescription.
The next step in the DFT/EFT program will be to develop systematic
expansions to these contributions in terms of $\rho$ and $\tau$. 

For now, we carry out the DFT/EFT formalism in the 
effective action framework using
the hybrid functional with $\tau(\xvec)$.
The full effective action is given by 
\beq
 \til{\Gamma}[\rho,\tau]=
 \til{\Gamma}_0[\rho,\tau]+\til{\Gamma}_1[\rho,\tau]+
\sum_{k=2}^{\infty}\,  \til{\Gamma}_k[\rho,\tau] 
\label{EffAction}
\ .
\eeq
We proceed to calculate the sources using Eqs.~(\ref{eq:finala}),
(\ref{eq:final}), and 
(\ref{eq:Correlation}) to NNLO; first $J_0(\xvec)$ is 
\bea
J_0(\xvec) 
  &=&  \frac{\delta}{\delta\rho(\xvec)}
     \left( \til{\Gamma}_1[\rho,\tau] + \sum_{k=2}^3
             \til{\Gamma}_k[\rho,\tau] \right)
  = -{\delta\over\delta\rho(\xvec)}
     \left( E_{\rm HF}[\rho]+ E_{c}[\rho,\tau]\right)
\nonumber \\
     &=&
 -\frac{(\nu-1)}{\nu}\,\frac{4\pi\, a_s}{M}\,\rho(\xvec)
     - 
  {7\over3}\, b_1\, \frac{a_s^2}{2M}\, [\rho(\xvec)]^{4/3} - 
  \frac{8}{3}\,b_4\, 
  \frac{ a_s^3}{2M}[\rho(\xvec)]^{5/3}
\nonumber \\
     & &
- \left(B_2\,a_s^2 \, r_s +B_3\,a_p^3\right)\frac{1}{2M}\,\tau(\xvec) 
+ \left(3B_2\,a_s^2 \, r_s
-B_3\,a_p^3\right)\frac{1}{4M}\,{\nabla}^2 \rho(\xvec)
\ ,     
      \label{eq:J0solve}
\eea  
and then $\eta_0(\xvec)$ is:
\beq
\eta_0(\xvec)
= -{\delta\over\delta\tau(\xvec)}
     \left( E_{\rm HF}[\rho]+ E_{c}[\rho,\tau]\right) 
     = - \left(B_2\,a_s^2 \, r_s
     +B_3\,a_p^3\right)\frac{1}{2M}\,\rho(\xvec)
\ .     
      \label{eq:eta0solve}
\eeq 
The spatially dependent effective mass $M^{*}(\xvec)$ is therefore
\beq
\frac{1}{2M^{*}(\xvec)} = \frac{1}{2M} - \eta_0(\xvec) 
= \frac{1}{2M} + \left[{(\nu-1)\over {4\nu}}\,C_2
     +{(\nu+1)\over {4\nu}}\,C'_2\right]\,\rho(\xvec)
     \ ,     
      \label{eq:MassEq2}
\eeq
where \Eq{C2imatch} has been used. 
An expression for the total binding energy (through NNLO)
follows by substituting for $J_0(\xvec)$ and $\eta_0(\xvec)$ 
in \Eq{G0} and then using \Eq{EffAction} and \Eq{eq:LamE},
\bea
\hspace*{-2.0 cm}  E[\rho(\xvec),\tau(\xvec)] 
   &=& \nu \sum_k^{\rm occ.}\varepsilon_k
   - \int\!\dthreex \,\biggl\{
     \frac{1}{2}\frac{(\nu-1)}{\nu}\,\frac{4\pi\, a_s}{M} \,[\rho(\xvec)]^2 
     + \frac{4}{3}\, b_1\, \frac{a_s^2}{2M} \,[\rho(\xvec)]^{7/3}
     \nonumber \\
  &+&\null 
     \frac{5}{3}\,b_4\, \frac{a_s^3}{2M} \,[\rho(\xvec)]^{8/3} 
     + \left(3B_2\,a_s^2 \, r_s - B_3\,a_p^3\right)
             \frac{1}{8M}\,{[\nabla \rho(\xvec)]}^2
    \nonumber \\ 
  &+&\null  
     \left(B_2\,a_s^2 \, r_s +B_3\,a_p^3\right)
      \frac{1}{2M}\,\rho(\xvec)\tau(\xvec)
  \biggr\}
     \label{eq:Econ}
     \ .
\eea

The noninteracting case
($C_i \equiv 0$) uses $J_0(\xvec)\equiv 0$, $\eta_0(\xvec)\equiv 0$ 
and the first term in
Eq.~(\ref{eq:Econ}).
Leading order (LO) uses the first term in Eq.~(\ref{eq:J0solve})
and the first two terms in Eq.~(\ref{eq:Econ}), and so on for NLO and
NNLO.
An alternative expression for the energy is obtained by using the
second part of \Eq{G0} followed by Eqs.~(\ref{T0}), (\ref{EffAction}), and
(\ref{eq:LamE}):
\bea
E[\rho(\xvec),\tau(\xvec)] &=& 
\int\!\dthreex\, \biggl\{
  {1\over 2M}\tau(\xvec) + \,v(\xvec)\, \rho(\xvec) + 
\frac{1}{2}\frac{(\nu-1)}{\nu}\,\frac{4\pi\, a_s}{M}
    \, [\rho(\xvec)]^2
\nonumber \\
     &+&
\left(B_2\,a_s^2 \, r_s
+B_3\,a_p^3\right)\frac{1}{2M}\,\rho(\xvec)\,\tau(\xvec)
+ \left(3B_2\,a_s^2 \, r_s
-B_3\,a_p^3\right)\frac{1}{8M}\,{[\nabla \rho(\xvec)]}^2     
\nonumber \\
     &+&
  b_1\, \frac{a_s^2}{2M}\,[\rho(\xvec)]^{7/3}
+ b_4\,  \frac{a_s^3}{2M} \, [\rho(\xvec)]^{8/3} 
 \biggr\}
\ .   
\label{eq:Econ1}
\eea

\subsection{Comparison to Skyrme Hartree-Fock}
\label{sec:skyrme}

A nucleus is a self-bound system, so the external potential
$v(\xvec)=0$.
To compare the DFT/EFT functional to the conventional Skyrme energy
density functional of \Eq{eq:SkyrmeEdens}, we set the
spin multiplicity $\nu=4$ and use 
\Eq{C2imatch} to rewrite \Eq{eq:Econ1} in terms of the $C_i$'s,
obtaining the energy density:
\bea
 {\cal E}[\rho(\xvec),\tau(\xvec)] 
  = {1\over 2M}\tau(\xvec) &+& \frac{3}{8}C_0\,[\rho(\xvec)]^2 + 
      \frac{1}{16}(3C_2 + 5C^{'}_2)\rho{(\xvec)}\,\tau{(\xvec)} 
     + \frac{1}{64}(9C_2 - 5C^{'}_2)(\bm{\nabla} \rho)^2
 \nonumber \\
     &+&
    b_1\, \frac{a_s^2}{2M}[\rho(\xvec)]^{7/3}
   + b_4\,  \frac{a_s^3}{2M}  [\rho(\xvec)]^{8/3} \ .   
\label{eq:SKConnect}
\eea
We observe that we get all the terms of $\edens_{SK}(\xvec)$
from \Eq{eq:SkyrmeEdens} except for the one with
coefficient $t_3$ and the spin-dependent terms, if we make the 
correspondence $t_0 \leftrightarrow C_0$, $t_1 \leftrightarrow
C_2$ and $t_2 \leftrightarrow C^{'}_2$. 
This correspondence is not surprising since the Skyrme interaction was
originally motivated as a low-momentum expansion of the G matrix
\cite{SKYRME}.
The two additional terms in \Eq{eq:SKConnect} of the form
${\rho(\xvec)}^{2+\alpha}$ come from correlations (i.e., terms beyond
Hartree-Fock), but there is no direct association with the $t_3$
term in the Skyrme energy density, which was originally motivated as a
three-body contribution (so $\alpha=1$).
However, it is clear that the Skyrme functional is incomplete as an
expansion; a direct connection to microscopic interactions by matching
to an EFT will include at least these additional terms. 

The generalization of the DFT/EFT to include spin and isospin dependence is
straightforward.  If one writes a complete set of four-fermion terms
with $\bm{\sigma}$ and $\bm{\tau}$ matrices in the EFT Lagrangian,
there are redundant terms because of Fermi statistics.
In conventional discussions of
the Skyrme approach, this observation 
is typically cast in terms of antisymmetrization
of the interaction \cite{VB72,RINGSCHUCK}.  
For a path integral formulation of the DFT/EFT, using Fierz
rearrangement is a convenient alternative.  
We illustrate the procedure for the leading spin dependence.

First consider just spin-1/2 (no isospin, so $\nu = 2$).
We expand the product of Grassmann fields $\psi_i \psi^\dagger_j$ 
($i$ and $j$ are spin indices) in the complete basis
of $\delta_{ij}$ and $\sigma^a_{ij}$ ($a=\{1,2,3\}$), identifying the
coefficients by contracting in turn with $\delta_{ij}$ and
$\sigma^b_{ji}$. The result, with minus signs from 
interchanging Grassmann fields, is
\beq
 \psi_i \psi^\dagger_j = -\frac{1}{2}(\psi^\dagger\psi)\delta_{ij}
   - \frac{1}{2}(\psi^\dagger \sigma^a \psi) \sigma^a_{ij}
   \ .
   \label{eq:fierza}
\eeq
If we substitute this result into
$(\psi^\dagger \psi)^2 = \psi^\dagger_i \{\psi_i \psi^\dagger_j\}
\psi_j$,
we find
\beq
  (\psi^\dagger \psi)^2 = -\frac12 (\psi^\dagger \psi)^2
    - \frac12 (\psi^\dagger \bm{\sigma} \psi)^2 \ ,
\eeq
or
\beq
  (\psi^\dagger \bm{\sigma} \psi)^2 = - 3 (\psi^\dagger \psi)^2
  \ .
\eeq
(We could also start with
 $(\psi^\dagger_i \sigma^a_{ii'} \psi_{i'}) 
 (\psi^\dagger_j \sigma^a_{jj'} \psi_{j'})$
and obtain the same result with a bit more effort).
Therefore
\beq
 C_0 (\psi^\dagger \psi)^2 + C_0^\sigma (\psi^\dagger \bm{\sigma} \psi)^2
  = (C_0 - 3C_0^\sigma) (\psi^\dagger \psi)^2
\eeq
and the single term $\til{C}_0 (\psi^\dagger \psi)^2$ with
$\til{C}_0 \equiv C_0 - 3 C_0^\sigma$ yields the
same results for all diagrams as the original two terms.

For the $\nu=4$ case with spin and isospin, we perform a
similar procedure to find
\beq
 \psi_{i\alpha} \psi^\dagger_{j\beta} = 
   -\frac{1}{4}(\psi^\dagger\psi)\delta_{ij}\delta_{\alpha\beta}
   - \frac{1}{4}(\psi^\dagger \sigma^a \psi) \sigma^a_{ij}\delta_{\alpha\beta}
   -\frac{1}{4}(\psi^\dagger \tau^b \psi)\delta_{ij}\tau^b_{\alpha\beta}
   -\frac{1}{4}(\psi^\dagger \sigma^a \tau^b \psi)
       \sigma^a_{ij}\tau^b_{\alpha\beta}   
   \ .
   \label{eq:fierzb} 
\eeq
Substituting into any two of $(\psi^\dagger \psi)^2$, 
$(\psi^\dagger \bm{\sigma} \psi)^2$,
$(\psi^\dagger  \bm{\tau}\psi)^2$, and
$(\psi^\dagger  \bm{\sigma} \bm{\tau}\psi)^2$, we find two independent
relations, which can be solved simultaneously to find:
\bea
  (\psi^\dagger \bm{\tau} \psi)^2 &=& - 2 (\psi^\dagger \psi)^2
    - (\psi^\dagger \bm{\sigma} \psi)^2
  \\
  (\psi^\dagger \bm{\sigma}\bm{\tau} \psi)^2 
     &=& - 3 (\psi^\dagger \psi)^2
     \ ,
\eea
which allow us to eliminate explicit dependence on $\bm{\tau}$
matrices in favor of just two independent couplings:
\bea
  &&
 C_0 (\psi^\dagger \psi)^2 + C_0^\sigma (\psi^\dagger \bm{\sigma} \psi)^2
 + C_0^\tau (\psi^\dagger  \bm{\tau}\psi)^2 
 + C_0^{\sigma\tau} (\psi^\dagger  \bm{\sigma} \bm{\tau}\psi)^2
  \nonumber \\
 && \qquad\qquad
  = (C_0 - 2C_0^\tau - 3C_0^{\sigma\tau}) (\psi^\dagger \psi)^2
    + (C_0^\sigma - C_0^\tau) (\psi^\dagger \bm{\sigma} \psi)^2
    \ .
\eea
This agrees (of course) with the usual discussion in terms of
antisymmetrized interactions (e.g., see
\cite{RINGSCHUCK} or \cite{VB72}).
The choice to eliminate the $\bm{\tau}$ terms is purely conventional.
The convention with Skyrme interactions is to choose the independent
couplings to be $t_i$ and $x_i$, which multiply terms in the
effective interaction in the combination
$t_i (1 + x_i P_\sigma)$, with $P_\sigma$ the spin exchange operator.
Extending to more derivatives and the spin-orbit (and tensor) terms
follows systematically in the EFT approach, but introduces many more
constants;
complete sets of contact terms with more derivatives
in the EFT Lagrangian can be found in Refs.~\cite{VANKOLCK} and
\cite{EPELBAUM}.
The proliferation of constants leads to a clash of philosophies between
the minimalist, phenomenological approach (use as few terms as
possible), which is necessarily model dependent,
and the model-independent EFT approach (use a complete set of terms).

The similarities of the successful Skyrme functional and the DFT
functional for a dilute Fermi gas prompts an analysis of typical Skyrme
parameters as effective range parameters. 
One can use the association of the $t_i$'s and the $C_i$'s along with
numerical values from successful Skyrme parameterizations
(e.g., Ref.~\cite{BROWN98})
to estimate
``equivalent'' values of $a_s$, $r_s$, and $a_p$.
We find that $a_s \approx -2\mbox{--}3\,\mbox{fm}$, 
which is about the inverse pion mass and is
much smaller than the large, fine-tuned values of
the free-space nucleon-nucleon interaction (however,
some Skyrme parameterization
have an ``equivalent'' $a_s$ of $5\,\mbox{fm}$ or larger).
However, $\kf a_s$ is still significantly larger than unity inside a
nucleus, which precludes a perturbative dilute expansion.  
Interestingly, the equivalent $r_s$ and $a_p$ values have magnitudes
consistent with what
one might expect from the nuclear hard-core radius (with $a_p < 0$), 
leading to $\kf r_s$
and $\kf a_p$ less than unity.

\section{Results for Dilute Fermi System in a Trap}
\label{sect:results}

In this section, we present numerical
results for the dilute Fermi system comprised of a small number 
of fermions confined in a harmonic oscillator trap.
We compare the nature of the convergence
of the EFT in a finite system both qualitatively and
quantitatively to the analysis done purely in the LDA \cite{PUG02}, 
which means the effect of treating the Hartree-Fock
contributions at NNLO [Fig.~\ref{fig:hug}(i) and (j)] exactly.
\subsection{Kohn-Sham Self-Consistent Procedure}

We restrict our calculations to finding the Kohn-Sham
orbitals for closed shells, so the density and potentials
are functions only of the radial coordinate $r \equiv |\xvec|$.
Note that the basic procedure is the \emph{same} one used for closed-shell
nuclei in Skyrme-Hartree-Fock \cite{RINGSCHUCK}, even though the DFT/EFT
can include correlations to any order.
The Kohn-Sham iteration procedure is as follows:
 \begin{enumerate}
   \item 
   Start by solving the Schr\"odinger equation with the 
   external potential profile $v(r)$ for the lowest $A$ states
   (including degeneracies) to find a set of
   orbitals and Kohn-Sham eigenvalues  
        $\{\psi_k,\varepsilon_k\}$. 
   
   \item
   Compute the density and kinetic energy density from the orbitals:
    \beq
      \rho(r) 
            = \sum_{k=1}^{A} |\psi_k({\bf x})|^2 \ ,
     \eeq
     \beq
      \tau(r) = \frac{1}{4\pi}\int\!d\Omega\,\tau(\xvec) 
 = \frac{1}{4\pi}\sum_{k=1}^{A}\int\!d\Omega\, |\nabla\,\psi_k({\bf
	    x})|^2 \ .
    \eeq

   \item 
   Using Eqs.~(\ref{eq:J0solve})--(\ref{eq:eta0solve}), find
   $J_0(r)$ and $\eta_0(r)$. 
   Evaluate the local single-particle potential 
   \beq
      v_s[\rho(r),\tau(r)] \equiv v_s(r) \equiv v(r) - J_0(r) 
        \label{eq:lspp}
   \eeq  
  at the chosen level of approximation (e.g., NLO) and the
  \lq\lq effective" mass :
  \beq
  \frac{1}{2M^{*}(r)} = \frac{1}{2M}- \eta_0(r)
  \ .
 \label{eq:effmass}
   \eeq 
    
   \item 
   Solve the Skyrme-type Schr\"odinger equation
   for the lowest $A$ states (including degeneracies), to find 
        $\{\psi_k,\varepsilon_k\}$ as before:
 \beq
  \bigl[ - \nabla \frac{1}{2M^{*}(r)}\nabla  +  v_s(r)
  \bigr]\, \psi_k({\bf x}) = \varepsilon_k \psi_k({\bf x})
  \, .
 \eeq

   \item
    Repeat steps 2.--4.\ until changes are acceptably
    small (``self-consistency'').  In practice, the changes in the
    density are ``damped'' by using a weighted average of
    the densities from the $(n-1)$th and $n$th iterations:
    \beq
       \rho(r) = \beta \rho_{n-1}(r) + (1-\beta) \rho_n(r)
       \ ,
    \eeq
    with $0 < \beta \le 1$.
  \end{enumerate}
  
This procedure has been implemented for dilute fermions in a trap using
two different methods for carrying out step~4. 
The Kohn-Sham single-particle equations are solved in one approach
by direct integration
of the differential equations 
and in the other approach 
by diagonalization 
of the single-particle Hamiltonian
in a truncated basis of unperturbed harmonic oscillator
wavefunctions.  The same results are obtained to high accuracy.
[Note that other methods used for Skyrme-type equations, such as the
conjugate-gradient method \cite{REINHARD}, can also be directly applied.]

\subsection{Fermions in a Harmonic Trap}

The interaction through NNLO is specified in terms of the three
effective range parameters $a_s$, $r_s$, and $a_p$. 
For the numerical calculations presented here, we consider the natural
case of hard-sphere 
repulsion with radius $R$, 
in which case $a_s=a_p=R$ and $r_s=2R/3$,
and also the case with $a_p = 2a_s$, so we can emphasize the
effect of $M^*(r)/M$ significantly less than unity.

Lengths are measured in units of the oscillator
parameter $b \equiv \sqrt{\hbar/M\omega}$, 
masses in terms of the fermion mass $M$, and
$\hbar = 1$.
In these units, $\hbar\omega$ for the oscillator is unity and
the Fermi energy of a non-interacting gas with
filled shells up to $N_F$ is $E_F = (N_F + 3/2)$.
The total number of fermions $A$ is related to $N_F$ by
\beq
   A = \frac{\nu}{6}(N_F+1)(N_F+2)(N_F+3) \ .
\eeq
Since we have only considered spin-independent interactions, 
our results are independent of whether the spin degeneracy $\nu$ actually
originates from spin, isospin, or some flavor index.

With interactions included, single-particle states are labeled by a
radial quantum number $n$, an orbital angular momentum $l$ with
$z$-component $m_l$, and the spin projection.  The radial functions
depend only on $n$ and $l$, so the degeneracy of each level
is $\nu\times(2l+1)$.
The solutions take the form (times a spinor, which is suppressed)
\beq
  \psi_{nlm_l} (\xvec) = R_{nl}(r)\, Y_{lm_l}(\Omega)
    = \frac{u_{nl}(r)}{r}\, Y_{lm_l}(\Omega)
   \ ,
\eeq
where the radial function $u_{nl}(r)$ satisfies
\beq
 \left[
   - \frac{1}{2 M^\ast(r)} \frac{d^2}{dr^2} 
   - \frac{d\eta_{0}}{dr}\,
     \left(\frac{1}{r} - \frac{d}{dr} \right)
   + v_s(r) 
   + \frac{l(l+1)}{2M^\ast(r) r^2} 
 \right] 
  u_{nl}(r) = \varepsilon_{nl} u_{nl}(r)
  \ ,
\eeq
and the $u_{nl}$'s are normalized according to
\beq
   \int_0^\infty \! |u_{nl}(r)|^2\, dr = 1 \ .
\eeq
Thus the density is given by
\beq
  \rho(r) 
     = \nu\, \sum_{nl}^{\rm occ.}\, \frac{(2l+1)}{4\pi}\,|R_{nl}(r)|^2 \ , 
\eeq
and the kinetic energy density is given by \cite{REINHARD}
\beq
\tau(r) =
\frac{\nu}{4\pi}\, \sum_{nl}^{\rm occ.}\,(2l+1)
\left[
{\left(
\frac{dR_{nl}}{dr}
\right)}^{2}
+ \frac{l(l+1)}{r^2}\,
 |R_{nl}(r)|^{2}
\right]
\ .
\eeq
The interactions are sufficiently weak that the occupied states are in
one-to-one correspondence with those occupied in the non-interacting
harmonic oscillator potential.

\subsection{Numerical Results}

\begin{figure}[p]
\centerline{\includegraphics*[width=13cm,angle=0]{DensityDistribution240}}
\vspace*{-.1in}
\caption{NNLO Kohn-Sham density distributions for a dilute gas of fermions in a harmonic
trap with degeneracy $\nu=2$ filled up to $N_{\rm F}=7$, which implies
there are 240 particles in the trap.  The scattering length is
$a_s = 0.16$ and the effective range is $r_s = 2a_s/3$.  Results for
two values of $a_p$ are compared for the LDA $\rho$-only functional 
($\rho$--DFT) and
with $\tau$ ($\rho\tau$--DFT).}
\label{fig:ks1}
%
%
%
\vspace*{.2in}

\centerline{\includegraphics*[width=13cm,angle=0]{DensityDistributionDiff240}}
\vspace*{-.1in}
\caption{Deviation of $\rho\tau$--DFT from $\rho$--DFT results at NNLO 
for the same systems
as in Fig.~\ref{fig:ks1}.}
\label{fig:ks2}
\end{figure}
\begin{figure}[t]
\centerline{\includegraphics*[width=13cm,angle=0]{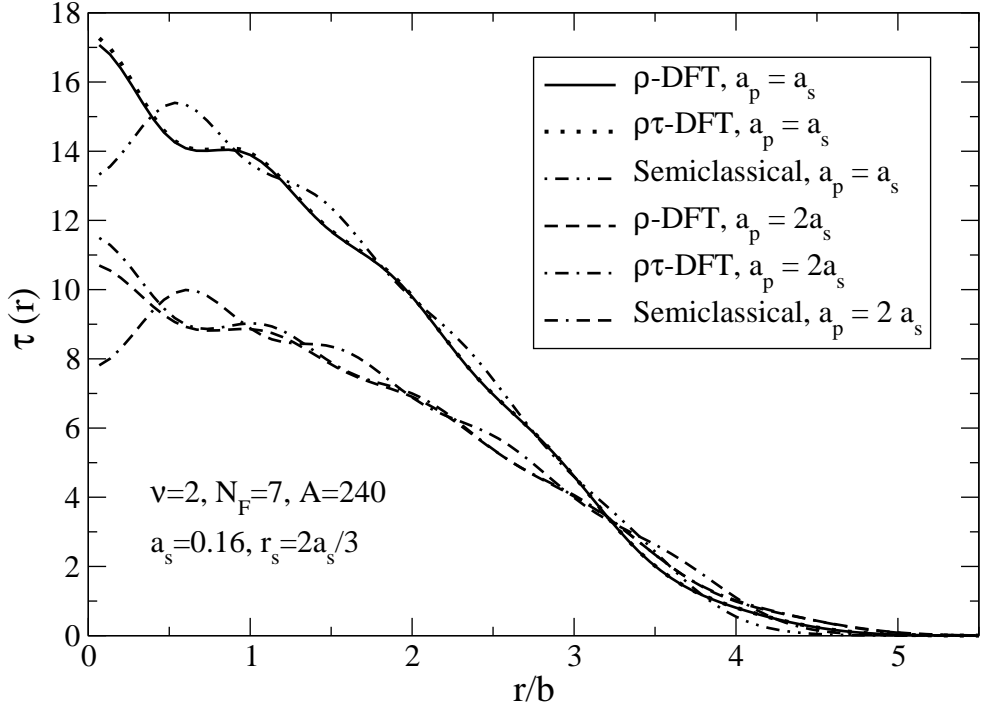}}
\vspace*{-.1in}
\caption{NNLO Kohn-Sham kinetic-energy-density 
distributions for the same systems as in Fig.~\ref{fig:ks1}.
The upper three curves are for $a_p=a_s$ and the lower three curves are
for $a_p=2a_s$.}
\label{fig:tau}
\end{figure}
\begin{figure}[t]
\centerline{\includegraphics*[width=14cm,angle=0]{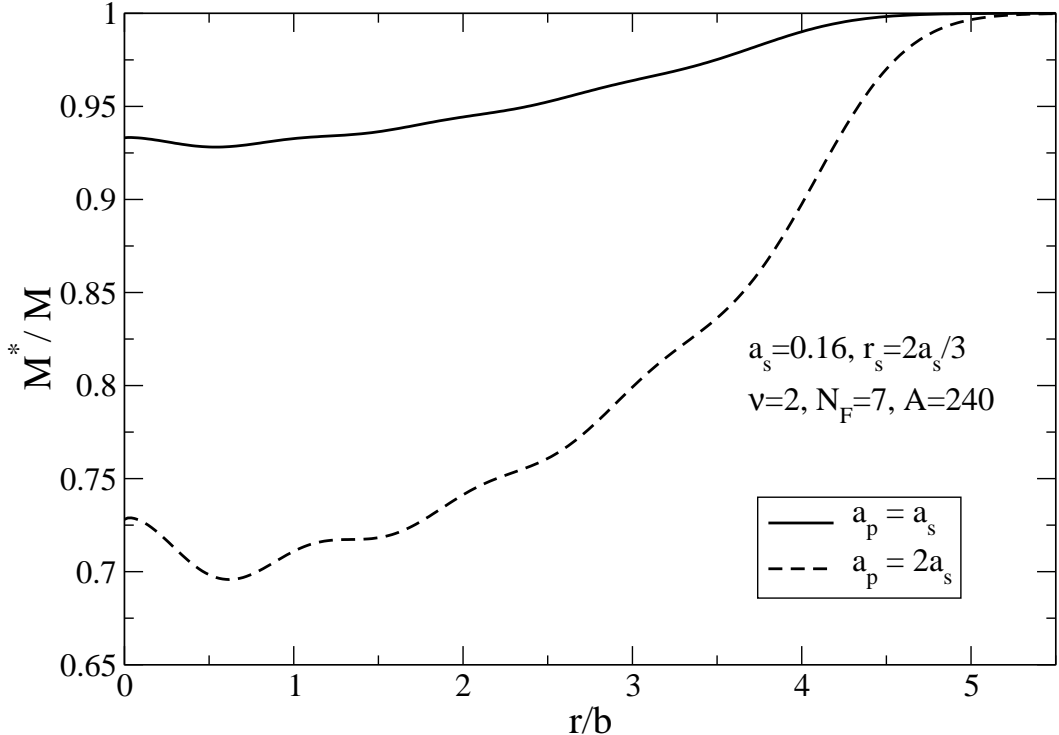}}
\vspace*{-.1in}
\caption{NNLO Kohn-Sham effective mass
distributions  for the same systems as in Fig.~\ref{fig:ks1}.
Results for two values of $a_p$ are shown.}
\label{fig:mstar}
\end{figure}
\begin{table}
\renewcommand{\tabcolsep}{12pt}
\caption{\label{tab:TrapResults}%
Energies per particle, averages
of the local Fermi momentum $\kf$, and rms radii for 
sample parameters and particle numbers 
for a dilute Fermi gas in a harmonic trap.  
See the text for a description of units.
The scattering length is fixed at $a_s = 0.16$ and the effective range is set
to $r_s = 2a_s/3$
when $a_p \neq 0$.  Results with the DFT functional including
$\tau$ are marked ``$\tau$--NNLO.''}
\begin{ruledtabular}
\begin{tabular}{ccrcdcdc}
  \multicolumn{1}{c}{$\nu$}  &  
  \multicolumn{1}{c}{$N_F$} &  
  \multicolumn{1}{c}{$A$} &  
  \multicolumn{1}{c}{$a_p$}   & 
  \multicolumn{1}{c}{$E/A$}  & 
  \multicolumn{1}{c}{$\langle \kf\rangle$} & 
  \multicolumn{1}{c}{$\sqrt{\langle r^2\rangle}$} & 
  approximation \\ \hline
  2 & 7 &  240 &   --  &  7.36  &  3.08  &  2.76  & LO \\
  2 & 7 &  240 &   --  &  7.51  &  3.03  &  2.81  & NLO (LDA) \\
  2 & 7 &  240 &   0.00  &  7.52  &  3.02  &  2.82  & NNLO (LDA) \\
  2 & 7 &  240 &   0.16  &  7.66  &  2.97  &  2.87  & NNLO (LDA) \\
  2 & 7 &  240 &   0.16  &  7.65  &  2.97  &  2.87  & $\tau$--NNLO (LDA) \\
  \hline
  2 & 7 &  240 &   0.32  &  8.33  &  2.76  &  3.10  & NNLO (LDA) \\
  2 & 7 &  240 &   0.32  &  8.30  &  2.77  &  3.09  & $\tau$--NNLO (LDA) \\
 \end{tabular}
\end{ruledtabular}
\end{table}
\begin{figure}[t]
\centerline{\includegraphics*[width=14cm,angle=0]{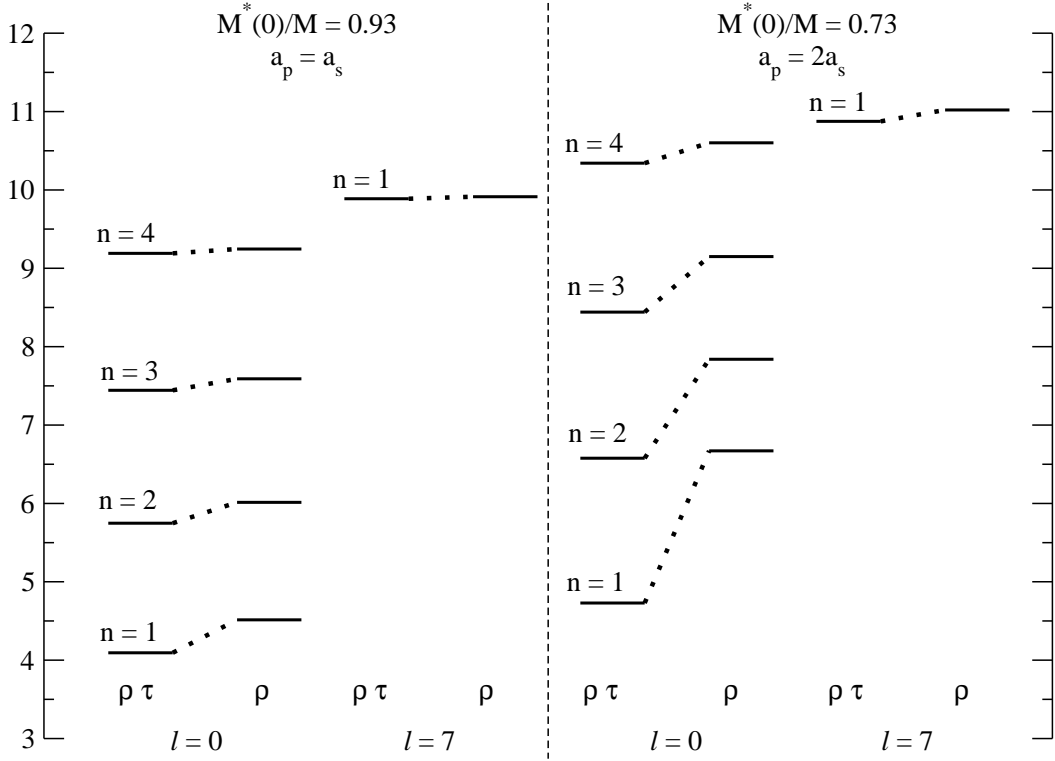}}
\vspace*{-.1in}
\caption{Comparison of selected single-particle energy spectra
 for the same systems as in Fig.~\ref{fig:ks1}.}
\label{fig:spectra}
\end{figure}

Here we compare
density distributions at zero temperature for the LDA analysis \cite{PUG02}
to those from evaluating 
Figs.~\ref{fig:hug}(i) and (j) exactly at NNLO (where the differences first
appear in the present analysis).
This comparison obviously
makes sense only if at least one of $r_s$ and $a_p$ is non-zero. 
We choose 240 particles as a representative example (other numbers of
particles give qualitatively similar results).
Densities at different orders in the DFT expansion were shown in
Ref.~\cite{PUG02}.  Results for the energy per particle $E/A$, average
Fermi momentum $\langle \kf\rangle$ and the rms radius are given in 
Table~\ref{tab:TrapResults}. 

In Fig.~\ref{fig:ks1}, we compare the densities at NNLO in the Kohn-Sham
formalism with $\rho$-only functionals (the LDA calculation from
Ref.~\cite{PUG02}) to the same system with the $\rho\tau$ functionals.
For hard-sphere scattering (for which $a_p = a_s$), the density curves
are almost indistinguishable.  If we plot the difference on an expanded
scale (see Fig.~\ref{fig:ks2}), we can see a small amplitude
oscillation.  The close agreement is not surprising given that the
source of the difference is the NNLO Hartree-Fock terms, so the
difference itself is higher order in the EFT
expansion
(note that the $\rho$--DFT and $\rho\tau$--DFT NNLO energies in
Table~\ref{tab:TrapResults} differ only by 0.01).  We can magnify the
difference by considering $a_p = 2a_s$, which multiplies the
corresponding Hartree-Fock term by a factor of eight (which also implies
that the $C'_2$ coefficient is unnaturally large).  For this case, the
difference in Fig.~\ref{fig:ks1} is visible and significant oscillations are
seen in Fig.~\ref{fig:ks2}.
The increased oscillation is analogous to the difference between
Thomas-Fermi and Kohn-Sham DFT densities (shown in Ref.~\cite{PUG02}),
although not as dramatic. 
The explanation is also analogous: the $\rho\tau$--DFT captures more
non-locality into the DFT functional.

The kinetic energy densities for these cases are shown in
Fig.~\ref{fig:tau}.
When calculated from the Kohn-Sham wave functions, $\tau(r)$ is quite
similar for the $\rho$-only and $\rho\tau$ calculations.
Also shown in this figure is the leading contribution from the
semiclassical approximation, which reproduces the ``exact'' kinetic
energy densities except near the origin. 

Only in the $\rho\tau$ case is the effective mass $M^*(r)$ different
from unity; it is shown in Fig.~\ref{fig:mstar} for the two values of
$a_p$.  The values close to the origin are in the range of those obtained in
Skyrme functionals fit to nuclear data.  In those functionals, the value
of $M^*$ is associated with the single-particle energy levels.  This
correspondence is seen in the single-particle energy spectra
in Fig.~\ref{fig:spectra}.  
If we take
the uniform limit (with the external potential turned off),
the single-particle energies for momentum ${\bf k}$ differ according to
\beq
   \varepsilon^{\rho}_{\bf k} - \varepsilon^{\rho\tau}_{\bf k}
     = (B_2 a_s^2 r_s + B_3 a_p^3) 
     \frac{(\kf^2 - {\bf k}^2)}{2 M} \rho
    \ .
\eeq
Thus, for positive $r_s$ and $a_p$, the $\rho\tau$ levels will always
lie lower except at the Fermi surface (where they must be equal).

This comparison 
demonstrates how the Kohn-Sham formalism can be misinterpreted.  Even
though the single-particle energies differ significantly, they are
not observables.  Indeed, the true bulk observables calculated in the
DFT framework, those in Table~\ref{tab:TrapResults}, are barely
distinguishable.  
One could ask
whether the single-particle levels in some
representation are ``better'' than in other representations.
In particular, they can be compared to  
the energy spectrum corresponding to the poles of the
exact Green's function, which can be constructed
in terms of the Kohn-Sham Green's function
\cite{VALIEV97b,FURNSTAHL04b}.
If this construction is carried out in the present approximations
for both the $\rho$-only and the $\rho\tau$ formalisms, the $\rho\tau$ 
single-particle spectrum is obtained in both cases \cite{FURNSTAHL04b}.
It is not clear from the present calculation how close the full and
Kohn-Sham spectra would be if the LDA were relaxed 
[e.g., for Fig.~\ref{fig:hug}(b)].
It would be useful to add spin-orbit interactions 
and then to study Kohn-Sham spin-orbit splittings near
the Fermi surface, since such splittings are
commonly fit in mean-field models.

\subsection{Power Counting and Convergence}
\label{sec:PCC}

The effective field theory approach
allows us to estimate contributions to the energy. 
At each successive order in the EFT expansion, the  low-energy constants
(LEC's) can be estimated using naive dimensional analysis, or NDA
\cite{PUG02}. 
In the case of a short-range force with a natural scattering length,
the underlying momentum scale $\Lambda \sim 1/R$, where $R$ is the range
of the potential, is the basic ingredient in the NDA. 
The estimate of two-body Hartree-Fock
energy contributions from a given term in the Lagrangian
can be found by replacing $\psi^\dagger\psi$ by the average density
(and including an appropriate spin factor) and the coefficient
by the natural estimate $C_{2i} \sim 4\pi/M\Lambda^{2i+1}$ \cite{HAMMER00}.   
As an example, the
NDA estimate of the Hartree-Fock energy per particle at LO was computed
from \Eq{EHF} as :
\beq
{\left(E_{\rm HF}\over A\right)}_{\rm NDA} 
    \approx {1\over2} {(\nu-1)\over \nu} \ \frac{4\pi}{M\Lambda} \
\langle\rho (\xvec)\rangle
\ ,
\eeq
with $\Lambda = 1/R = 1/a_s$ for a hard-sphere potential. 
The Thomas-Fermi result can be used to find the average density, 
or one might just take the actual computed average value (the results
will differ by much less than the uncertainty in the estimate).
The NLO estimate was found by multiplying the LO
result by $\langle\kf\rangle a_s$, where $\langle\kf\rangle$ is the
average Fermi momentum \cite{PUG02}. At NNLO, we have three terms.
The ${\rho}^{8/3}$ LDA term was estimated by multiplying the LO
result by$(\langle\kf\rangle a_s)^2$, and the other two terms 
($\rho\,\tau \mbox{ and } \nabla \rho$) arising from Hartree-Fock at
that order was estimated directly as in the case of LO (using
$\langle\rho\tau\rangle \approx \langle\rho\rangle \langle\tau\rangle$). 

\begin{figure}[p]
\centerline{\includegraphics*[width=9.7cm,angle=0]{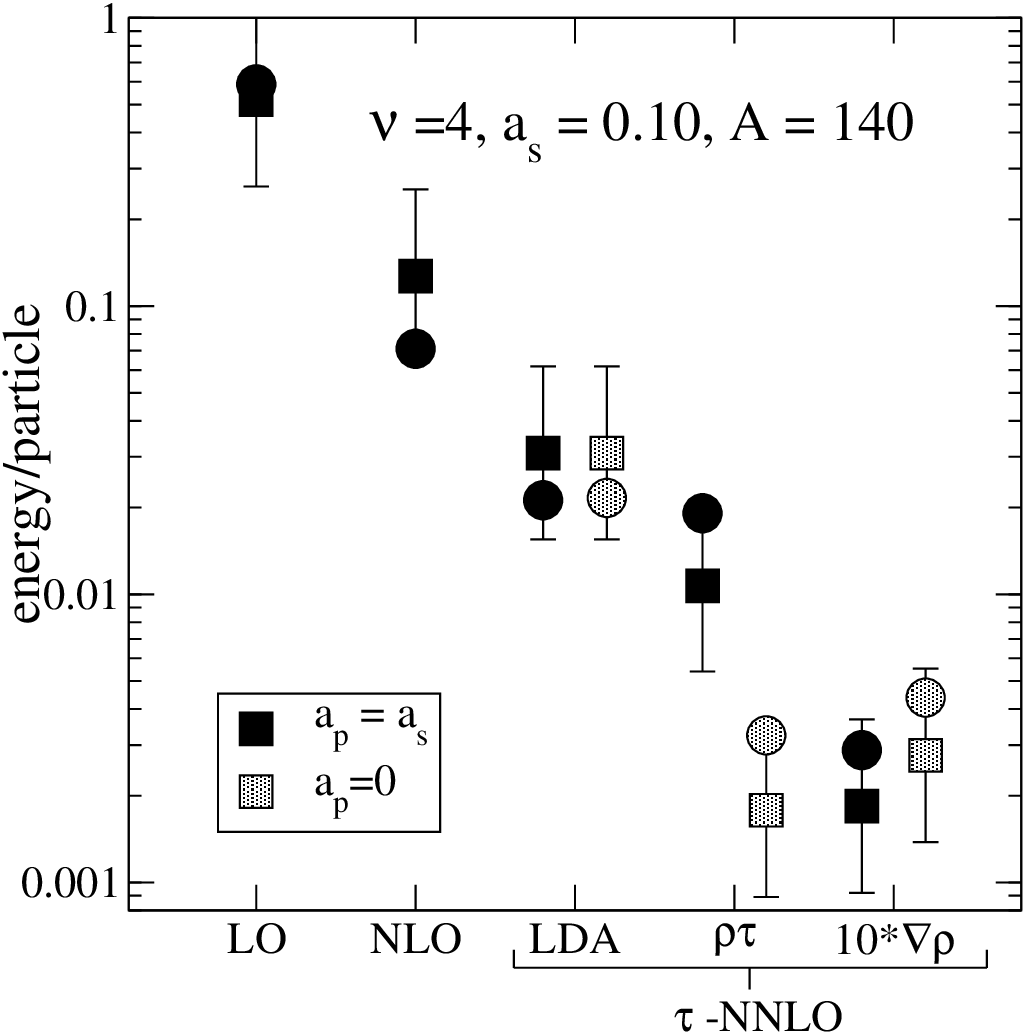}}
\vspace*{-.1in}
\caption{Energy estimates (squares and error bars) 
for $\nu=4$, $a_s = 0.10$, $A = 140$ particles
compared with actual
values (circles) for a hard-sphere gas (solid) and with the p-wave
scattering length equal to zero (shaded).}
\label{fig:errplt}
%
%
\vspace*{.1in}
%
\centerline{\includegraphics*[width=9.7cm,angle=0]{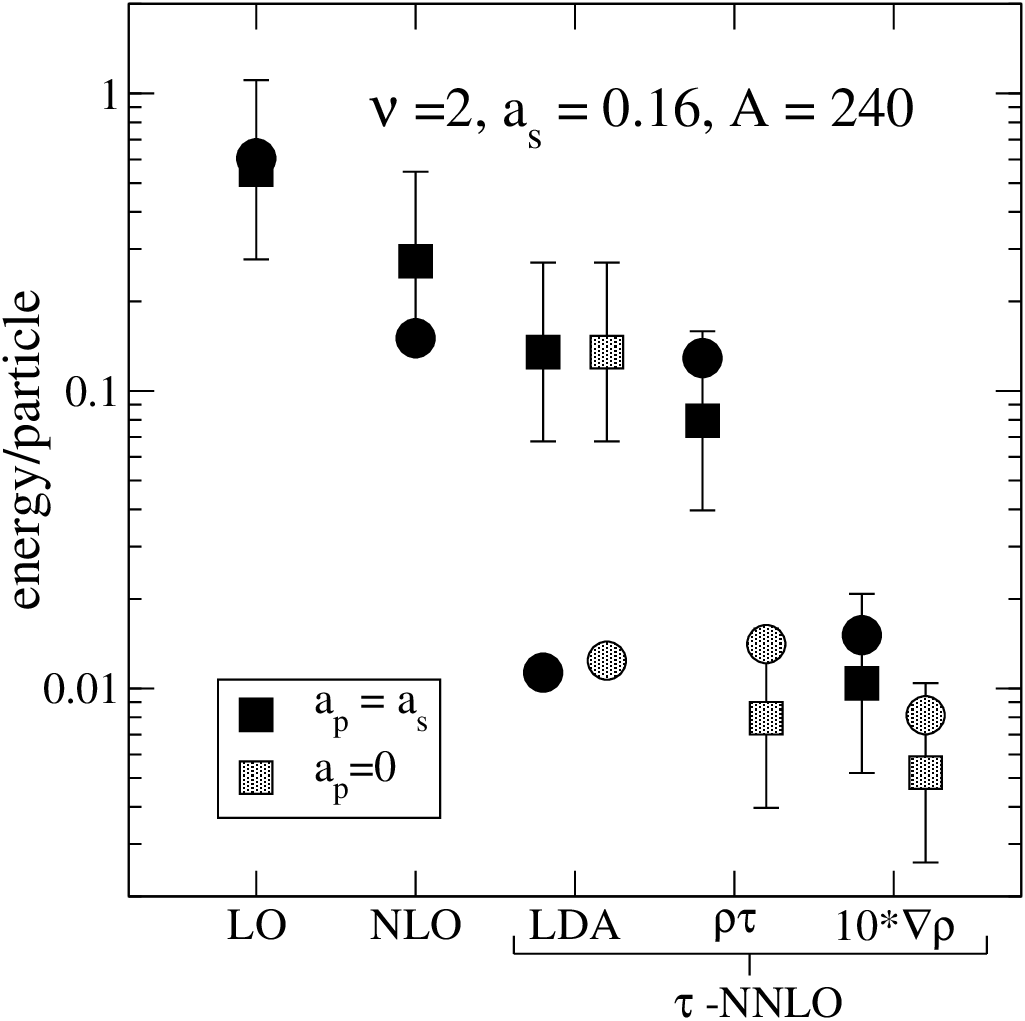}}
\vspace*{-.1in}
\caption{Energy estimates (squares and error bars) 
for $\nu=2$, $a_s = 0.16$, $A = 240$ particles
compared with actual
values (circles) for a hard-sphere gas (solid) and with the p-wave
scattering length equal to zero (shaded).}
\label{fig:errplt2}
\end{figure}

In Figs.~\ref{fig:errplt} and \ref{fig:errplt2}, 
estimates and actual contributions are shown for $\nu=4$,
$A=140$, $a_s = 0.10$
(for which $\langle \kf a_s\rangle \approx 0.24$) and for
$\nu=2$, $A=240$, $a_s = 0.16$ 
(for which $\langle \kf a_s\rangle \approx 0.5$).
Square symbols denote estimates based on naive 
dimensional analysis,
with error bars indicating a 1/2 to 2 uncertainty in the
estimate. Actual contributions to the energy per particle from each of
the orders are shown as round symbols, i.e., 
the actual NLO contribution is  
$|E_{\rm NLO}-E_{\rm LO}|/A$. 
At NNLO, we plot estimated contributions from the
${\rho}^{8/3}$ LDA, $\rho\,\tau$, and  gradient terms separately. 
The latter gives a very small contribution consistent with
its NDA estimate, and it has been multiplied by ten in
Figs.~\ref{fig:errplt} and \ref{fig:errplt2} 
to fit them on the graphs.
The two sets of estimates and results correspond to $a_p = a_s$
(hard sphere) and $a_p = 0$.

We see from Figs.~\ref{fig:errplt} and \ref{fig:errplt2} 
that the actual results for LO,
NLO and NNLO agree 
well with NDA estimates, including the new gradient contributions, 
with one exception. 
The exception is the $\rho^{8/3}$ LDA estimate for the $\nu=2$ system, which
greatly overestimates the actual contribution at that order due to 
an accidental cancellation when $\nu=2$
between the two terms in the 
$b_4$ coefficient in Eq.~(\ref{eq:bfour}) \cite{PUG02}. 
In general, however, we can use these estimates to reliably predict
the uncertainty in the energy per particle from higher orders.

While it's clear that nuclei are \emph{not} perturbative dilute Fermi
systems
with natural free-space scattering lengths, there is phenomenological
evidence that power counting can apply to energy functionals that are
fit to bulk nuclear properties. 
Phenomenologically successful functionals, both of the Skyrme type
and covariant, have Hartree terms that are consistent with
Georgi-Manohar NDA for a chiral low-energy theory 
\cite{Furnstahl:2001hs,FURNSTAHL99b,HACKWORTH}.
This involves power counting with two scales: the pion decay constant
$f_\pi$ and an underlying scale for short-range physics $\Lambda$, which
empirically (for these functionals) is around 600\,MeV.
(Equivalently, there is an additional large dimensionless coupling
$g \sim \Lambda/f_\pi$ that enters in a well-prescribed manner).
Thus the functionals \emph{do} take the form of a density expansion
(with parameter $\rho/f_\pi^2\Lambda$) as
well as a gradient expansion, with the same hierarchy as illustrated
here.
A major goal of future investigations will be to elucidate the nature
of the density expansion for finite nuclei and to connect it to the
underlying chiral EFT.

%

%
\section{Summary}
\label{sect:summary}

In this paper,
the EFT-based Kohn-Sham density functional for a confined, dilute Fermi
gas was extended
by incorporating the kinetic-energy-density $\tau(\xvec)$ 
into the formalism.
The generating functional is constructed by including, in addition to a source 
$J(\xvec)$ coupled
to the composite density operator $\psi^\dagger\psi$, 
another source $\eta(\xvec)$ coupled to the (semi-local) kinetic energy
density operator $\nabla\psi^\dagger\cdot\nabla\psi$.   
A functional Legendre transformation with respect to the sources
yields an effective action of the kinetic
energy density $\tau$ as well as the fermion density $\rho$. 
This construction also serves as a prototype for including additional
densities currents (such as separate proton and neutron densities
or a spin-orbit current).

This extension sets the stage for the construction of energy functionals
and Kohn-Sham equations that go systematically
beyond the local density approximation (LDA).
As a first step, we included
the exact Hartree-Fock (HF) contribution at NNLO for a natural
dilute Fermi system but treated non-HF contributions
in LDA.
This exact HF contribution provides explicit dependence 
on $\tau$ and on the gradient of the density, unlike the LDA.
The EFT expansion for a confined, finite system with natural effective
range parameters
is controlled by those parameters
($a_s$, $r_s$, $a_p$, \ldots) times an average Fermi momentum
and by the gradient of the density.  
Thus the expansion is perturbative in the sense of being a density
expansion but is not perturbative in an underlying potential.
(This is clear since our prototype system is hard spheres, which yields
infinities even in first-order perturbation theory!)
An error plot of contributions to the energy per particle versus the
order of the calculation showed that we can reliably estimate the
truncation error in a finite system, including the gradient
terms. 

The ground-state 
energy functional and Kohn-Sham single-particle equations
constructed here take the same form
as those in Skyrme Hartree-Fock calculations (not including the
spin-orbit contribution, which can be added with a similar
generalization). 
There is work supporting the Skyrme (and also covariant
\cite{FURNSTAHL99b}) mean-field
energy functionals as density expansions.  Low-energy effective
theories of QCD are expected to have a type of power counting.
As shown in Ref.~\cite{HACKWORTH}, power counting in the energy
functional based on chiral naive dimensional analysis is fully
consistent with phenomenological parametrizations.
The value of including the kinetic energy density as a functional
variable is not conclusive in the present example, 
but it does lead to a single-particle spectrum closer to that of the
exact Green's function.
In general,
summing more nonlocalities without decreasing the numerical efficiency of
the calculation is a plus.

Even in the simplest low-density expansion, there are contributions
at all powers of the Fermi momentum, which means fractional powers
of the density.
In the Skyrme parametrization, there is only one such term. 
This is presumably a balance between phenomenological accuracy and a
desire to maximize predictive power through limiting parameters.
For reproducing bulk properties of stable nuclei,
the Skyrme functional is only sensitive to a relatively small
window in density,
which allows significant freedom for a phenomenological
parameterization.
The future challenge will be to see if additional terms motivated by
power counting, which may become important for extrapolating
far from stability, can be determined.

The immediate
next steps are to develop a derivative expansion to evaluate
diagrams such as the NLO ``beachball'' diagram beyond LDA and to test it for
convergence, and to generalize the DFT formalism to include 
pairing. 
Work is in progress on 
these extensions, which will be directly relevant for 
nuclear applications \cite{FURNSTAHL04}.
In addition, in order
to adapt the density functional procedure to chiral effective field
theories with explicit pions, 
we will need to extend the discussion to include long-range
forces. 
Kaiser, Weise, and collaborators have already generated functionals of the
Skyrme type using free-space chiral perturbation theory together with
the density matrix expansion of Negele et al. \cite{Kaiser:2002jz}.  
It is not clear that the
power counting is consistent in those calculations, 
but future development along these lines is certainly warranted.
Finally, a systematic solution to 
the large scattering length problem for trapped atoms also remains a
challenge.

\acknowledgments

We thank A.~Bulgac, J.~Dobaczewski, J.~Engel, H.-W.~Hammer, 
W.~ Nazarewicz, S.~Puglia,
A.~Schwenk, and B.~Serot for useful comments and discussions.
This work was supported in part by the National Science Foundation
under Grant No.~PHY--0098645.


\end{document}